\documentclass[a4paper]{JHEP3}
\usepackage{latexsym,amssymb}
\usepackage{graphicx}

\usepackage{braket}
\usepackage{nicefrac}
\usepackage{comment}
\usepackage[latin1]{inputenc}
\usepackage{amsmath}
\usepackage{amscd}
\usepackage{Macros}
\usepackage{exscale}
\usepackage[symbol*]{footmisc}

\renewcommand{\uni}[1]{\ensuremath{1\!\!1_{#1 × #1}}}
\title{Heterotic strings on homogeneous spaces\thanks{Research partially
    supported by the EEC under the contracts MEXT-CT-2003-509661,
    MRTN-CT-2004-005104, MRTN-CT-2004-503369 and HPRN-CT-2000-00122.}  }

\author{ Dan Israël${}^{\bigstar \spadesuit}$, Costas Kounnas${}^\Diamond$, Domenico
  Orlando${}^\spadesuit$, P. Marios Petropoulos${}^\spadesuit$\\

  \begin{itemize}
    
  \item   Racah Institute of Physics, The Hebrew University \\
    Jerusalem 91904, Israel
    
  \item Laboratoire de Physique Théorique de l'École Normale
    Sup{é}rieure\footnote{Unit{é} mixte du CNRS et de l'École Normale
      Sup{é}rieure, UMR 8549.} \\
    24 rue Lhomond, 75231 Paris Cedex 05, France
    
  \item   Centre de Physique Théorique, École Polytechnique\footnote{Unité
    mixte du CNRS et de l'École Polytechnique, UMR 7644.} \\
  91128 Palaiseau, France
  \end{itemize}

\bigskip

E-mail: \email{israeld@phys.huji.ac.il}, \email{kounnas@lpt.ens.fr},
\email{orlando@cpht.polytechnique.fr}, \email{marios@cpht.polytechnique.fr}
}

\abstract{We construct heterotic string backgrounds corresponding to 
families of homogeneous spaces as exact conformal field theories. 
They contain left cosets of compact groups by their maximal tori  
supported by \textsc{ns-ns} 2-forms and gauge field fluxes. We
give the general formalism and modular-invariant partition functions, then
we consider some examples such as $SU \left( 2 \right) / U \left( 1
\right) \sim S^2 $ (already described in a previous paper) and the $SU
\left( 3\right)/ U\left( 1 \right)^2 $ flag space. As an application 
we construct new supersymmetric string vacua with magnetic fluxes 
and a linear dilaton.}

\preprint{
LPTENS-0439 \\
CPHT-RR052.0904\\bi
hep-th/0412220}


\begin{document}


%

\section{Motivations and summary}

The search for exact string backgrounds has been a major motivation in the
field for many years. Gravitational backgrounds with a clear geometric
interpretation are even more important since they may provide a handle on
quantum gravitational phenomena, black holes and ultimately cosmology -- for
those which are time-dependent. Wess--Zumino--Witten models provide such a
class of solutions, with remarkable properties. The target space is in that
case a group manifold and, together with the metric, the Neveu--Schwarz
antisymmetric tensor is the only background field. Both of these fields are
exactly known to all orders in $\alpha'$. So are the spectrum, partition
function, two- and three-point functions,~\dots

Wess--Zumino--Witten models appear in many physical set-ups, as near-horizon
geometries of specific brane configurations. The three-sphere is part of the
near-horizon geometry of $N_5$ NS5-branes. This is the target space of an
$SU(2)_k$ super-\textsc{wzw} model at bosonic 
level $k= N_5 -2 $. Another celebrated
example is that of AdS$_3$.  The latter appears in the
NS5-brane/fundamental-string background, together with $S^3$, at equal
radius $L = \sqrt{\alpha' N_5}$; it is realized in terms of the $SL(2,
\mathbb{R})_{\tilde{k}}$ \textsc{wzw} at level $\tilde{k} = k + 4$. These
are important examples because of their role in the study of decoupling
limits, little-string theory, holographic dualities etc. The knowledge of
exact spectra, amplitudes, \dots is crucial for better understanding of
these issues.

Despite the many assets of \textsc{wzw} models, the major limitation comes
from the dimension and signature of their target spaces. When dealing with
compact groups, the dimension often exceeds six (e.g.  $SU(3)$ is
eight-dimensional), while for non-compact groups, $SL(2,
\mathbb{R})_{\tilde{k}}$ is the only example with a single time direction.

In order to reduce the dimension of the target space, while keeping
two-dimensional conformal invariance and tractability, the usual procedure
is the gauging. Gauged \textsc{wzw} models are realized algebraically, at
the level of the chiral currents and energy--momentum tensor, by following
the \textsc{gko} construction~\cite{Gliozzi:1976qd}. Alternatively, one can work
directly on the action and gauge symmetrically a subgroup $H \subset G$. For
$H = U(1)$, the gauged model can even be obtained as an extreme marginal
deformation of the original model, driven by a $\int \mathrm{d}^2 z J \bar J
$ perturbation, where $J$ and $\bar J$ are the currents associated with the
$U(1)\subset G$.

Target spaces of gauged \textsc{wzw} models \emph{are not} usual geometric
cosets $G/H$. Firstly, the background fields of gauged \textsc{wzw} receive
non-trivial $\alpha'$ corrections\footnote{The higher-order $\alpha'$
  corrections are trivial for \textsc{wzw} models: they boil down to
  shifting $k\to k + g^\star$ in the classical backgrounds ($g^\star$ is the
  dual Coxeter number of the group $G$).}, while geometric cosets can be
assigned a well-defined metric. Secondly, the isometry groups are different.
For geometric cosets, the isometry group is $G$, while it is $H$ for the
target space of the gauged \textsc{wzw}.

Geometric cosets could provide alternative backgrounds, with different
properties and new possibilities for accommodating six or less space
dimensions, or a single time direction (in the non-compact case).
Unfortunately, they have not been systematically analyzed, and were even
thought to be, at most, leading-order solutions to the string equations.
Although some exact solutions were identified in the past
\cite{Johnson:1995kv,Berglund:1996dv,Ferrara:1995ih}, no generic pattern for
generalization was known.

The issue of geometric cosets as exact backgrounds has been recently
revisited in~\cite{Israel:2004vv}. There, it was shown that $S^2 \equiv
SU(2)/U(1)$ and AdS$_2 \equiv SL(2,\mathbb{R})/U(1)_\textrm{space}$, with
magnetic and electric fluxes and no dilaton, can be obtained as extreme
marginal deformations of the $SU(2)_k$ and $SL(2,\mathbb{R})_{\tilde k}$
\textsc{wzw} models. In this case the background fields are exact up to the
usual finite renormalization of the radius ($k \to k + 2$ and $\tilde{k} \to
\tilde{k} - 2$) and spectra, partition functions, \dots are within reach.
The marginal deformations are \emph{asymmetric} because the right current
that appears in the bilinear does not belong to the right-moving affine
algebra of the group at hand.

Asymmetric marginal deformations apply to any group. The aim of the present
paper is to investigate on several interesting generalizations of this
method, in the case of compact groups, and make contact with
\emph{asymmetrically gauged} \textsc{wzw} models. We will focus in
particular on the $SU \left( 3 \right)$ group. In this case the asymmetric
marginal deformation leads to the $SU(3)/U(1)$ geometric coset, with
magnetic fluxes and no dilaton. In turn, this coset is identified with the
asymmetric gauging of a $U(1)^2$ in the original \textsc{wzw} models.

In the cases under consideration, however, more possibilities exist, which
we further exploit. We examine the asymmetric gauging of the full Cartan
torus $U(1)^2$. The geometric cosets obtained in this way, can be assigned
two different metrics depending on the precise manner the gauging is
performed, in combination with the extreme asymmetric marginal deformation.
One is K\"ahlerian and consequently no NS form survives: we obtain the flag
space $F_3 = SU(3)/U(1)^2$, recognized many years ago~\cite{Prati:1985us} 
to be a leading-order solution, thanks to its K\"ahlerian structure. %
The other metric is not K\"ahlerian, and the background has both magnetic and
NS fluxes. It enters into the construction of non-compact 
manifolds of G$_2$ holonomy~\cite{Atiyah:2001qf}.

All our solutions are exact sting backgrounds with no dilaton -- contrary to
the usual symmetrically gauged \textsc{wzw} models. We can determine their
spectra as well as their full partition functions. 

The paper is organized as follows: first we fix the notation by reviewing
some known facts about \textsc{wzw} models and then show how to read the
background fields corresponding to an asymmetric marginal deformation of
such models. We emphasize in particular the decompactification of the Cartan
torus that takes place at the extremal points in moduli
space (Sec.~\ref{sec:geom-constr}). This formalism is then used to study the
deformation of the $SU \left( 2 \right)$ and $SU \left( 3 \right)$ models
(Sec.~\ref{sec:some-examples}). In the following we introduce a different
construction in which the limit deformations are identified to
asymmetrically gauged \textsc{wzw} models~\cite{Quella:2002fk} and the
deformation is generalized so to reach the different constant-curvature
structures admitted by an asymmetric $G/T$ coset, with particular emphasis
on the $SU \left( 3 \right)/ U\left( 1 \right)^2 $ case
(Sec.~\ref{sec:gauging}). The next section
(Sec.~\ref{sec:modular-invariance}) deals with the computation of the
one-loop partition functions for the asymmetric deformations leading to
geometric cosets. Two different methods are proposed, one using the
Kazama--Suzuki decomposition in terms of Hermitian symmetric spaces, the
other via the direct deformation of the Cartan lattice of the Lie algebra
corresponding to the group. In the final section
(Sec.~\ref{sec:heter-strings-sing}) we give an example of application by
using these \textsc{scft}'s to construct other supersymmetric exact string
backgrounds such as the left-coset analogues of the NS5-branes solutions
\cite{Callan:1991dj,Kounnas:1990ud}. They provide new 
holographic backgrounds of the Little String Theory 
type~\cite{Aharony:1998ub,Giveon:1999px,Giveon:1999zm}, and may 
be dual to non-trivial supersymmetric compactifications on 
manifolds with singularities. The concluding appendices contain some
facts about the geometry of coset spaces, partition functions and characters
of affine Lie algebras.



\newcommand{\mJ}{\ensuremath{\mathcal{J}}}
\newcommand{\mI}{\ensuremath{\mathcal{I}}}
\newcommand{\mK}{\ensuremath{\mathcal{K}}}

\newcommand{\F}[2]{{f^{#1}_{\phantom{#1}#2}}}
\setcounter{footnote}{0}
%

\section{Compact coset spaces: general formalism}
\label{sec:geom-constr}

In this section we will fix the notation by reviewing some well known facts
about conformal field theories on group manifolds (\textsc{wzw} models) and
give the general formalism for the truly marginal deformations leading to
exact \textsc{cft}'s on left coset spaces.

\subsection{String theory on group manifolds: a reminder}
\label{sec:asymm-marg-deform}

Let $\mathfrak{g}$ be the (semi-simple) Lie algebra of the (compact) group
$\mathcal{G}$ and $\Set{ T_{\textsc{m}}}$ a set generators that satisfy the
usual commutation relations $\left[ T_{\textsc{m}}, T_{\textsc{n}} \right]
=\sum_{\textsc{p}} \F{\textsc{mn}}{\textsc{p}} T_{\textsc{p}}$ and are
normalized with respect to the Killing product $\kappa \left(
  T_{\textsc{m}}, T_{\textsc{n}} \right) = -\tr \left( T_{\textsc{m}}
  T_{\textsc{n}} \right) = \delta_{\textsc{mn}}$. We can always write
$\mathfrak{g}$ as the direct sum $\mathfrak{g} = \mathfrak{j} \oplus \mathfrak{k}$ where
$\mathfrak{k}$ is the Cartan subalgebra and correspondingly distinguish between
the Cartan generators $\Set{T_{a}}$ and the generators of $\mathfrak{j}$,
$\Set{T_{\mu }}$.

The generators are in one-to-one correspondence with the 
Maurer--Cartan left-invariant one-forms defined by:
\begin{equation}
\label{eq:Maurer-Cartan}
  \mathcal{J}_{\textsc{m}} = \kappa \left( T_{\textsc{m}}, g^{-1}\di g \right) =
- \tr \left( T_{\textsc{m}} g^{-1} \di g \right)
\end{equation}
where $g$ is the general element of the group $\mathcal{G}$. It is a well
known fact that the scalar product on $\mathfrak{g}$ naturally induces a scalar
product $\braket{\cdot,\cdot}$ on the tangent space $T_g$ to $\mathcal{G}$ that
can be written by decomposing the induced metric (the so-called
Cartan--Killing metric) in terms of the currents as follows:
\begin{equation}
  \label{eq:carkil}
  \braket{\di g, \di g } = \kappa \left( g^{-1} \di g_, 
    g^{-1} \di g \right) = \sum_{\textsc{mn}} \delta_{\textsc{mn}} \kappa
  \left( T_{\textsc{m}}, g^{-1} \di g \right) 
  \kappa \left( T_{\textsc{n}}, g^{-1} \di g \right) =  
  \sum_{\textsc{mn}} \delta_{mn} \mJ^{\textsc{m}} 
  \otimes \mJ^{\textsc{m}}
\end{equation}


Now let us consider the affine extension of the Lie algebra
$\hat{\mathfrak{g}}_k$, at level $k$. We have two sets of holomorphic and
anti-holomorphic currents of dimension one, naturally related to the 
 Maurer--Cartan right- and left-invariant one-forms:
\begin{equation}
\label{eq:holo-currents}
  J_{\textsc{m}} (z) = -k \ \kappa (T_{\textsc{m}} , \d g \, g^{-1}) \ , \ \ 
  \bar{J}_{\textsc{m}} (\bar z ) = k \ \kappa (T_{\textsc{m}} , g^{-1} \db g ).
\end{equation}
Each set satisfies the following operator product expansion:
\begin{equation}
  J_{\textsc{m}} (z) J_{\textsc{n}} (w) = \frac{k\delta_{\textsc{mn}}}{2(z-w)^2} 
  + \frac{\F{\textsc{mn}}{\textsc{p}}\, J_{\textsc{p}} (w)}{z-w} \ 
  + \ \mathcal{O} \left((z-w)^0 \right)
\end{equation}
This chiral algebra contains the Virasoro operator, given by the Sugawara
construction:
\begin{equation}
  T (z) = \sum_{\textsc{m}} \frac{:J_{\textsc{m}} J_{\textsc{m}}:}{k+g^*}
\end{equation}
where $g^{\ast}$ is the dual Coxeter number and the corresponding central
charge is given by:
\begin{equation}
  c = \frac{k \, \dim (\mathfrak g)}{k+g^\ast}.
\end{equation}
An $N=1$ superconformal extension is obtained by adding $\left(
  \dim \mathfrak{g} \right)$ free fermions transforming in the adjoint
representation:
\begin{align}
  T (z) &= \sum_{\textsc{m}} \frac{:J_{\textsc{m}} J_{\textsc{m}}:}{k+g^{\ast}} + : \psi_{\textsc{m}} \d \psi_{\textsc{m}}: \\
  G (z) &= \frac{2}{k} \left( \sum_{\textsc{m}} J_{\textsc{m}} \psi_{\textsc{m}} - \frac{i}{3k} 
   \sum_{\textsc{mnp}} f_{\textsc{mnp}} : \psi_{\textsc{m}} \psi_{\textsc{n}} \psi_{\textsc{p}} : \right) 
\end{align} 
An heterotic model is provided by considering a left-moving $N=1$
current algebra and a right-moving $N=0$ one.  The Lagrangian
($\sigma$-model) description of this model is given by the linear combination of
the following \textsc{wzw}-model and the action for free fermions
transforming in the adjoint representation:
\begin{equation}
  S = \frac{k}{4\pi} \int_{\d \mathcal{B}} \Tr \left( 
    g^{-1} \di g \land \ast \ g^{-1} \di g \right) 
  + \frac{k}{12\pi} \int_{\mathcal{B}} \Tr \left(
    g^{-1} \di g \right)^3
  + \frac{1}{2\pi} \int d^2 z \ \psi_{\textsc{m}} \db \psi_{\textsc{m}}
\end{equation}
(the exterior derivative is here understood as acting on the worldsheet
coordinates). The background fields corresponding to this action are the
Cartan-Killing metric Eq.~\eqref{eq:carkil} and the \textsc{ns-ns} two-form
field, coming from the \textsc{wz} term:
\begin{equation}
  H = \di B = \Tr (g^{-1} \di g )^3 = \frac{1}{2} f_{\textsc{mnp}}\  
  \mathcal{J}^{\textsc{m}} \land \mathcal{J}^{\textsc{n}} \land \mathcal{J}^{\textsc{p}}
\end{equation}

\subsection{Asymmetric deformations}
\label{sec:asymm-deform}

Truly marginal deformations of \textsc{wzw} models were already studied in
\cite{Chaudhuri:1989qb,Forste:2003km}. In particular in heterotic 
strings we can consider a
deformation obtained with the following exactly marginal operator $V$ built from the
total Cartan currents of $\mathfrak{g}$ (so that it preserves the local
$N=(1,0)$ superconformal symmetry of the theory):
\begin{equation}
  \label{margdef}
  V = \frac{\sqrt{kk_g}}{2\pi} \int \di^2 z \  \sum_{a} \textsc{h}_{a} 
  \left( J^{a} (z) - \frac{i}{k} \F{a}{\textsc{mn}} : 
    \psi^{\textsc{m}} \psi^{\textsc{n}} : \right) \bar J (\bar{z} )
\end{equation}
(where the set $\{ \textsc{h}_{a}\} $ are the parameters of the deformation 
and $\bar J (\bar z )$ is a right moving
current of the Cartan subalgebra of the heterotic gauge group 
at level $k_g$). Such a deformation is always truly marginal since the $J_a $
currents commute.


It is not completely trivial to read off the deformed background fields that
correspond to the $S+V$ deformed action. A possible way is a method
involving a Kaluza--Klein reduction as in~\cite{Horowitz:1995rf}. For
simplicity we will consider the bosonic string with vanishing dilaton and
just one operator in the Cartan subalgebra $\mathfrak{k}$. The right-moving
gauge current $\bar J$ used for the deformation has now a left-moving
partner and can hence be bosonized as $\bar J = \imath \bar \partial \varphi $, $\varphi \left(
  z, \bar z \right) $ being interpreted as an internal degree of freedom.
The sigma-model action is recast as
\begin{equation}
  \label{eq:KK-action}
  S = \frac{1}{2 \pi} \int \di^2 z \: \left( G_{\textsc{mn}} + B_{\textsc{mn}} \right)
  \partial x^{\textsc{m}} \bar \partial x^{\textsc{n}},
\end{equation}
where the $x^{\textsc{m}}, \textsc{m}=1,\ldots,4$ embrace the group coordinates
$x^\mu, \mu = 1,2,3$ and the internal $x^4 \equiv \varphi$:
\begin{equation}
  x^{\textsc{m}} = \left( \begin{tabular}{c}
      $  x^\mu $\\ \hline
      $ x^4$
    \end{tabular}\right).
\end{equation}
If we split accordingly the background fields, we obtain the following
decomposition:
\begin{align}
  G_{\textsc{mn}} = \left( \begin{tabular}{c|c}
      $G_{\mu\nu } $& $ G_{\varphi \varphi } A_\mu $\\ \hline
      $ G_{\varphi \varphi } A_\mu $& $G_{\varphi \varphi }$
    \end{tabular}\right), &&
  B_{\textsc{m}\textsc{n}} =  \left( \begin{tabular}{c|c}
      $B_{\mu\nu}$ & $B_{\mu 4}$ \\ \hline
      $-B_{\mu 4}$ & 0
    \end{tabular}\right),
\end{align}
and the action becomes:
\begin{multline}
  S = \frac{1}{2 \pi} \int \di z^2 \left\{ \left( G_{\mu \nu } + B_{\mu \nu} \right) \d x^\mu \db x^\nu
    + \left( G_{\varphi \varphi } A_\mu + B_{\mu 4} \right) \d x^\mu \db \varphi \right.\\
  \left. + \left( G_{\varphi \varphi } A_\mu - B_{\mu 4} \right)
    \d \varphi \db x^\mu + G_{\varphi \varphi } \d \varphi \db
    \varphi\right\}.
\end{multline}
  
We would like to put the previous expression in such a form that space--time
gauge invariance,
\begin{align}
  A_\mu & \to A_\mu + \d_\mu \lambda,  \\
  B_{\mu 4} & \to B_{\mu 4} + \d_\mu \eta,
\end{align}
is manifest. This is achieved as follows:
\begin{multline}
  S = \frac{1}{2 \pi } \int \di^2 z \: \left\{\left( \hat G_{\mu \nu } + B_{\mu
        \nu }\right) \d x^\mu \db x^\nu + B_{\mu 4} \left( \partial x^\mu \bar \partial \varphi
      - \partial \varphi \bar \partial x^\mu \right) \right. +\\ \left. + G_{\varphi \varphi } \left( \partial
      \varphi + A_\mu \partial x^\mu \right) \left( \bar \partial \varphi + A_\mu \bar \partial x^\mu
    \right)\right\},
\end{multline}
where $\hat G_{\mu \nu }$ is the Kaluza--Klein metric
\begin{equation}
  \hat G_{\mu \nu } = G_{\mu \nu } - G_{\varphi \varphi } A_{\mu } A_{\nu    }.
\end{equation}
We can then make the following identifications:
\begin{subequations}
  \label{eq:KK-fields}
  \begin{align}
    \hat G_{\mu \nu } &=  \frac{k}{2} \left( \mJ_\mu \mJ_\nu - 2 \textsc{h}^2
      \tilde{\mJ}_\mu \tilde{\mJ}_\nu \right) \label{eq:KK-metric}\\
    B_{\mu \nu } &= \frac{k}{2} J_\mu \land J_\nu ,     \label{eq:B-field} \\
    B_{\mu 4} &= G_{\varphi \varphi } A_\mu  =  \textsc{h} \sqrt{\frac{k k_g}{2}} \tilde{\mJ}_\mu, \\
    A_{\mu} &=  \textsc{h} \sqrt{\frac{2k}{k_g}} \tilde{\mJ}_\mu, \label{eq:KK-em-field} \\
    G_{\varphi \varphi } & = \frac{k_g}{2}.
  \end{align}
\end{subequations}

Let us now consider separately the background fields we obtained so to give
a clear geometric interpretation of the deformation, in particular in
correspondence of what we will find to be the maximal value for the
deformation parameters $\textsc{h}_a$.


\paragraph{The metric.}

According to Eq.~\eqref{eq:KK-metric}, in terms of the target space metric,
the effect of this perturbation amounts to inducing a back-reaction that in
the basis of Eq.~\eqref{eq:carkil} is written as:
\begin{equation}
\label{eq:def-metric}
  \braket{\di g, \di g}_{\textsc{h}} = \sum_{M} \mJ_M \otimes \mJ_M - 2 \sum_{a}
  \textsc{h}_{a}^{2} \mJ_a \otimes \mJ_a = \sum_\mu \mJ_\mu \otimes \mJ_\mu + \sum_a \left( 1 - 2 \textsc{h}_a^2
  \right) \mJ_a \otimes \mJ_a
\end{equation}
where we have explicitly separated the Cartan generators. From this form of
the deformed metric we see that there is a ``natural'' maximal value $\textsc{h}_a =
1/ \sqrt{2}$ where the contribution of the $\mJ_a \otimes \mJ_a $ term changes
its sign and the signature of the metric is thus changed. One could naively
think that the maximal value $\textsc{h}_a= 1/ \sqrt{2}$ can't be attained since the
this would correspond to a degenerate manifold of lower dimension; what
actually happens is that the deformation selects the the maximal torus that
decouples in the $\textsc{h}_{a} = \textsc{h} \to 1/ \sqrt{2}$ limit as it was shown in
\cite{Israel:2004vv} for the $SU \left( 2\right)$ and $SL \left( 2, \setR
\right)$ algebras.

To begin, write the general element $g \in \mathcal{G} $ as $g = h t $ where $h \in
\mathcal{G}/ \mathcal{T}, t \in \mathcal{T}$. Substituting this decomposition
in the expression above we find:
\begin{multline}
  \braket{\di \left( h t\right), \di \left( h t\right)}_{\textsc{h}} = \tr \left(
    \left( h t\right)^{-1} \di \left( h t\right) \left( h t\right)^{-1} \di
    \left( h t\right) \right) - \sum_a 2 \textsc{h}_a^2 \tr \left( T_a
    \left( h t \right)^{-1} d \left( h t\right)^{-1}\right)^{ 2} = \\
  =  \tr \left( h^{-1} \di h h^{-1} \di h\right) + 2 \tr \left( \di
      t \ t^{-1} h^{-1} \di h \right)+ \tr \left( t^{-1} \di t\ t^{-1} \di
      t\right)+\\ - \sum_{a} 2 \textsc{h}_a^2 \left( \tr \left( T_a t^{-1}
      h^{-1} \di h\ t \right) + \tr \left( T_a t^{-1} \di
      t\right)\right)^{ 2}
\end{multline}
let us introduce a coordinate system $\left( \gamma_\mu , \psi_a \right)$ such as
the element in $\mathcal{G}/ \mathcal{T}$ is parametrized as $h = h\left(
  \gamma_\mu \right)$ and $t$ is written explicitly as:
\begin{equation}
  t = \exp \left\{ \sum_{a} \psi_{a} T_{a} \right\}  =  
\prod_{a} e^{ \psi_{a} T_{a}}    
\end{equation}
it is easy to see that since all the $T_a $ commute $t^{-1} \di t = \di t\ 
t^{-1} = \sum_a T_a \di \psi_a $. This allows for more simplifications in the
above expression that becomes:
\begin{multline}
  \braket{\di \left( h t\right), \di \left( h t\right)}_{\textsc{h}} = \tr \left(
    h^{-1} \di h h^{-1} \di h\right) + 2 \sum_a \tr \left( T_a h^{-1} \di h
  \right)  \di \psi_a + \sum_a \di \psi_a  \di \psi_a +\\
- \sum_{a} 2 \textsc{h}_a^2\left( \tr
    \left( T_a h^{-1} \di h \right) + \di \psi_a\right)^{ 2} =  \tr \left(
    h^{-1} \di h h^{-1} \di h\right) - \sum_{a} 2 \textsc{h}_a^2\left( \tr \left( T_a
      h^{-1} \di h \right) \right)^{ 2}+\\ + 2  
\sum_a \left( 1 - 2\textsc{h}_a^2 \right) \tr
  \left( T_a h^{-1} \di h \right)   \di \psi_a + \sum_a \left( 1 - 2
    \textsc{h}_a^2\right) \di \psi_a  \di \psi_a
\end{multline}
if we reparametrise the $\psi_a $ variables as:
\begin{equation}
  \psi_a = \frac{\hat \psi_a }{\sqrt{1-2 \textsc{h}_a}}  
\label{rescale}
\end{equation}
we get a new metric $\braket{\cdot, \cdot}_\textsc{h}^\prime $ where we're free to 
take the $\textsc{h}_a \to 1/ \sqrt{2}$ limit:
\begin{multline}
  \braket{\di \left( h t\right), \di \left( h t\right)}_{\textsc{h}}^\prime = \tr \left(
    h^{-1} \di h h^{-1} \di h\right) - \sum_{a} 2 \textsc{h}_a^2\left( \tr \left( T_a
      h^{-1} \di h \right) \right)^{ 2} +\\ +2  
\sum_a \sqrt{ 1 - 2\textsc{h}_a^2} \tr
  \left( T_a h^{-1} \di h \right)  \di \hat \psi_a + 
\sum_a \di \hat \psi_a   \di \hat \psi_a
\end{multline}
and get:
\begin{equation}
\label{eq:G/TxT-metric}
  \braket{\di \left( h t\right), \di 
\left( h t\right)}_{\nicefrac{1}{ \sqrt{2}}}^\prime =
  \left[ \tr \left(
    h^{-1} \di h h^{-1} \di h\right) - \sum_{a} \left( \tr \left( T_a
      h^{-1} \di h \right) \right)^{  2} \right] + 
\sum_a \di \psi_a  \di \psi_a
\end{equation}
where we can see the sum of the restriction of the Cartan-Killing
metric\footnote{This always is a left-invariant metric on $G/H$. A symmetric
  coset doesn't admit any other metric. For a more complete discussion see
  App.~\ref{sec:coset-space-geometry}.} on $T_h \mathcal{G}/ \mathcal{T}$
and the metric on $T_t\mathcal{T} = T_tU\left( 1\right)^r$. In other words
the coupling terms between the elements $h \in \mathcal{G}/\mathcal{T}$ and
$t \in \mathcal{T}$ vanished and the resulting metric
$\braket{\cdot,\cdot}^\prime_{\nicefrac{1}{ \sqrt{2}}} $ describes the
tangent space $T_{ht}$ to the manifold $\mathcal{G}/\mathcal{T} ×
\mathcal{T}$.

  These homogeneous manifolds enjoy many interesting properties. The best
  part of them can be interpreted as consequence of the presence of an
  underlying structure that allows to recast all the geometric problems in
  Lie algebraic terms (see App.~\ref{sec:coset-space-geometry} for some
  constructions).  There's however at least one intrinsically geometric
  property that it is worth to emphasize since it will have many profound
  implications in the following. All these spaces can be naturally endowed
  with complex structures by using positive and negative roots as
  holomorphic and anti-holomorphic generators. Moreover for each space there
  is not in general only one of these structures (but for the lowest
  dimensional $SU \left( 2 \right)$ case) and there always exists one of
  them which is Kähler \cite{Borel:1958ch}.

\paragraph{Other Background fields.}
\label{sec:backgound-fields}

The asymmetric deformation of Eq.~(\ref{margdef}) generates a non-trivial
field strength for the gauge field, that from Eq.~\eqref{eq:KK-em-field} is
found to be:
\begin{equation}
  \label{eq:def-F-field}
  F^a = \sum_a \sqrt{\frac{2k}{k_g}} \textsc{h}_a \, \di \,\mathcal{J}^a = -
  \sum_a \sqrt{\frac{k}{2 k_g}} \textsc{h}_a  f^a_{\phantom{a} \mu\nu } J^\mu \land J^\nu 
\end{equation}
(no summation implied over $a$).\\
On the other hand, the $B$-field~\eqref{eq:B-field} is not changed, but the
physical object is now the 3-form $H$:
\begin{equation}
  \label{eq:def-H-field}
  H_{[3]} = \di B - \frac{1}{k_g} A^a \land \di A^a 
  = \frac{1}{2} f_{\textsc{mnp}} 
  \mathcal{J}^{\textsc{m}} \land \mathcal{J}^{\textsc{n}} \land \mathcal{J}^{\textsc{p}}  
  - 2 \sum_a  \textsc{h}_{a}^2 \, f_{a \textsc{np}} \,
  \mathcal{J}^a \land \mathcal{J}^{\textsc{n}} \land \mathcal{J}^{\textsc{p}} ,
\end{equation}
where we have used the Maurer-Cartan structure equations.  At the point
where the fibration trivializes, $\textsc{h}_a = 1/\sqrt{2}$, we are left with:
\begin{equation}
  H_{[3]} = \frac{1}{2} f_{\mu \nu \rho}\, 
  \mathcal{J}^{\mu} \land \mathcal{J}^{\nu} \land \mathcal{J}^{\rho}.
\end{equation}
So only the components of $H_{[3]}$ ``living'' in the coset $\mathcal{G}/
\mathcal{T}$ survive the deformation. They are not affected of course by the
rescaling of the coordinates on $\mathcal{T}$.

\paragraph{A trivial fibration.}
\label{sec:trivial-fibration}

The whole construction can be reinterpreted in terms of fibration as
follows. The maximal torus $\mathcal{T}$ is a closed Lie subgroup of the Lie
group $\mathcal{G}$, hence we can see $\mathcal{G}$ as a principal bundle
with fiber space $\mathcal{T}$ and base space $\mathcal{G}/ \mathcal{T}$
\cite{Nakahara:1990th}
\begin{equation}
  \mathcal{G} \xrightarrow{\mathcal{T}} \mathcal{G}/\mathcal{T}  
\end{equation}
The effect of the deformation consists then in changing the fiber and the
limit value $\textsc{h}_a = 1/ \sqrt{2}$ marks the point where the fibration becomes
trivial and it is interpreted in terms of a gauge field whose strength is
given by the canonical connection on $\mathcal{G}/ \mathcal{T}$
\cite{Kobayashi:1969}.

\subsection{Equations of motion}
\label{sec:equations-motion}

  In this section we want to explicitly show that the background fields we
  found on the left coset space are solution to the first order (in $\alpha^\prime$)
  equations of motion \cite{Callan:1985ia}. 
  
  For a vanishing dilaton they read:
  \begin{subequations}
    \label{eq:eqns-motion}
    \begin{align}
      \delta c &= -R + \frac{k_g}{16} F^a_{\phantom{a}\mu \nu}
      {F^{a}}^{\mu\nu}
      \label{eq:beta-phi} \\
      {\beta^{\left(G \right)}}_{\mu \nu} &= R_{\mu \nu} - \frac{1}{4} H_{\mu \rho \sigma}
      H_{\nu}^{\phantom{\nu} \rho\sigma} - \frac{k_g}{4} {F^a}_{\mu \rho}
      {F^a}_\nu^{\phantom{\nu }\rho} = 0\label{eq:beta-g} \\
      {\beta^{\left( B \right)}}_{\nu \rho} & = \nabla^\mu H_{\mu \nu \rho} = 0 \label{eq:beta-B}\\
      {\beta^{\left( A \right)}}_\mu & = \nabla^\nu {F^a}_{\mu \nu} - \frac{1}{2} {F^a}^{\nu
        \rho} H_{\mu \nu \rho} = 0 \label{eq:beta-A}
    \end{align}
  \end{subequations}
  after applying the proper normalizations\footnote{Unless explicitly stated
    we consider $\alpha^\prime = 1$ and the highest root $\psi = 2$} our
  fields are given by:
  \begin{subequations}
    \label{eq:bkd-fields}
    \begin{align}
      g_{\mu \nu } &= \frac{k}{2} \delta_{\mu \nu} \\
      {F^a}_{\mu \nu } &= - \sqrt{\frac{2k}{k_g}} f^a_{\phantom{a} \mu\nu } \\
      H_{\mu \nu \rho } &= - \frac{k}{2} f_{\mu \nu \rho }
    \end{align}
  \end{subequations}
  \begin{itemize}
  \item the $\beta^{\left( B \right)} = 0$ equation \eqref{eq:beta-B} is just
    the restriction of the same equation for the initial \textsc{wzw} model
  \item the two terms in the $\beta^{\left( A\right)} = 0 $ equation
    \eqref{eq:beta-A} vanish separately: the first one because $F$ is closed
    (or, equivalently because $\F{a}{\mu \nu }$ seen as a two form in $G/T$
    satisfies the condition stated below Eq.~\eqref{eq:two-form}); the second
    because it is proportional to:
    \begin{equation}
      \sum_{\nu, \rho \in \mathfrak{g/h}} f_{a\nu\rho} f_{\mu \nu \rho} = 
      \sum_{\textsc{m,r} \in \mathfrak{g}} f_{a\textsc{mr}} f_{\mu {\textsc{mr}}} = 2g^\ast \delta_{a\mu } = 0
    \end{equation}
  \item to solve the $\beta^{\left( G\right)} = 0 $ equation \eqref{eq:beta-g}
    we need some more work. Using the results in
    App.~\ref{sec:expl-deriv-some}, for a general algebra, we obtain
    \begin{equation}
      R_{\mu \nu } = \frac{1}{4}  \sum_{\rho, \sigma}f_{\mu \rho \sigma }
      f_{\nu \rho \sigma} + \sum_{a,\rho} f_{a \mu \rho } f_{a \nu \rho } 
    \end{equation}
    that is consistent with the result in Eq.\eqref{eq:coset-Ricci}.  
    
    If we introduce the orthonormal basis described in
    \eqref{eq:orthonormal-basis} the Ricci tensor can be explicitly written
    as:
    \begin{equation}
      R_{\mu \nu } = \frac{1}{4}  \sum_{\rho,\sigma}f_{\mu \rho \sigma }
      f_{\nu \rho \sigma} + \sum_{a,\rho} f_{a \mu \rho } f_{a \nu \rho } 
      = \frac{1}{2} g^\ast \delta_{\mu \nu} + \frac{1}{2}  \delta_{\mu \nu }
      \begin{cases}
        \abs{\alpha_{\nicefrac{(\mu + 1)}{2}}}^2
        & \text{if $\mu$ is odd}\\
        \abs{\alpha_{\nicefrac{\mu }{2}}}
        & \text{if $\mu$ is even}\\
      \end{cases}
    \end{equation}
    In particular for a simply laced algebra reduces to
    \begin{equation}
      R_{\mu \nu } = \frac{g^\ast + 2}{2} \delta_{\mu \nu} =
      \frac{g^\ast + 2}{k} g_{\mu \nu}   
    \end{equation}
    This result can be read by saying that the metric we obtain on a simply
    laced algebra is Einstein with the following Ricci scalar:
    \begin{equation}
      R =  \frac{g^\ast + 2}{k} \left( \dim \mathfrak{g} - \dim \mathfrak{k} \right) 
    \end{equation}
    For example in the case of $G = SU \left( N \right)$, $g^\ast = N$,
    $\left( \dim \mathfrak{g} - \dim \mathfrak{k} \right) = N \left( N - 1\right)$
    and then
    \begin{equation}
      R = \frac{\left( N + 2 \right) N \left( N -1\right)}{k} 
    \end{equation}
  \end{itemize}


\setcounter{footnote}{0}
%

\section{Some Examples}
\label{sec:some-examples}

In this section we will give some explicit examples of our construction. In
particular we will consider the deformation leading from the $SU \left( 2
\right)$ background to the $SU \left( 2 \right) / U \left( 1 \right) \sim S^2
$ coset (which already appeared in \cite{Israel:2004vv} as part of the
$\mathrm{AdS}_2 × S^2$ background) and 
the superconformal field theory on $SU \left( 3 \right) / U \left( 1 \right)^2 $. 
Although our construction is quite general and can in principle be applied
to any group there is a limited number of examples giving 
critical heterotic string theory backgrounds with a clear geometrical meaning.
This is just because of dimensional
reasons: $SU \left( 2 \right) / U \left( 1 \right)$ is two-dimensional, $SU
\left( 3 \right) / U \left( 1 \right)^2 $ is 6-dimensional and $USp\left( 4
\right) / U \left( 1 \right)^2 $ is 8-dimensional; higher groups on the
other hand would lead to cosets of dimension greater than 10 (in example $SU
\left( 4 \right) / U \left( 1 \right)^3 $ has dimension $15-3 = 12$). 
On the other hand these higher-dimensional cosets can be used e.g. 
to obtain non-trivial compactifications generalizing the 
constructions of~\cite{Gepner:1988qi,Kazama:1989qp}, if the level 
of these \textsc{cft}s are kept small.

\subsection{The two-sphere CFT}
\label{sec:two-sphere-cft}

The first deformation that we explicitly consider is the marginal
deformation of the $SU \left( 2 \right)$ \textsc{wzw} model. This was first
obtained in \cite{Kiritsis:1995iu} that we will closely follow. It is anyway
worth to stress that in their analysis the authors didn't study the point of
maximal deformation (which was nevertheless identified as a
decompactification boundary) that we will here show to correspond to the
2-sphere $S^2 \sim SU \left( 2 \right) \slash U \left( 1 \right)$. Exact
\textsc{cft}'s on this background have already obtained in
\cite{Bachas:1993kq} and in \cite{Johnson:1995kv}. In particular the
technique used in the latter, namely the asymmetric gauging of a $SU \left(
  2 \right) × U \left( 1 \right)$ \textsc{wzw} model, bears many
resemblances to our own.

Consider an heterotic string background containing the $SU(2)$ group
manifold, times some $(1,0)$ superconformal field theory $\mathcal{M}$.  The
sigma model action is:
\begin{equation}
  S = k S_{SU\left( 2 \right)} (g) + 
  \frac{1}{2\pi} \int d^2 z \ \left\{ 
    \sum_{a=1}^{3} \lambda^a \db \lambda^a
    + \sum_{n=1}^{g} \tilde{\chi}^n \d \tilde{\chi}^n 
  \right\} + S( \mathcal{M} ), 
\end{equation}
where $\lambda^i$ are the left-moving free fermions superpartners of the bosonic
$SU(2)$ currents, $\tilde{\chi}^n$ are the right-moving fermions of the
current algebra and $kS_{SU\left( 2 \right)}(g)$ is the \textsc{wzw} action
for the bosonic $SU(2)$ at level $k$. This theory possesses an explicit
$SU(2)_L × SU(2)_R$ current algebra.

A parametrization of the $SU \left( 2 \right)$ group that is particularly
well suited for our purposes is obtained via the so-called Gauss
decomposition that we will later generalize to higher groups (see 
App~\ref{sec:suleft-3right}). A general element $g \left( z, \psi \right) \in
SU \left( 2 \right)$ where $z \in \setC $ and $\psi \in \setR $ can be written as:
\begin{equation}
  g \left( z, \psi \right) = 
  \begin{pmatrix}
    1 & 0 \\ z & 1
  \end{pmatrix}
  \begin{pmatrix}
    1/\sqrt{f} & 0 \\ 0 & \sqrt{f}
  \end{pmatrix}
  \begin{pmatrix}
    1 & \bar w \\ 0 & 1
  \end{pmatrix}
  \begin{pmatrix}
    e^{\imath \nicefrac{\psi}{2}} & 0 \\ 0 & e^{- \imath \nicefrac{\psi}{2}}
  \end{pmatrix}
\end{equation}
where $w = - z $ and $f = 1 + \abs{z}^2$. In this parametrisation the
matrix of invariant one-forms $\Omega = g \left( z, \psi \right)^{-1} \di g
\left( z, \psi \right)$ appearing in the expression for the Maurer-Cartan
one-forms~\eqref{eq:Maurer-Cartan} is:
\begin{align}
  \Omega_{11} &= \frac{\bar z \di z - z \di \bar z + \imath f \di \psi }{ 2 f} &
  \Omega_{12} &= - \frac{e^{- \imath \psi }}{f} \di \bar z \\
  \Omega_{21} &= - \bar \Omega_{12} & \Omega_{22} &= - \Omega_{11}
\end{align}
(remark that $\Omega$ is traceless and anti-Hermitian since it lives in
$\mathfrak{su} \left( 2 \right)$). From $\Omega $ we can easily derive the
Cartan--Killing metric on $T_g SU \left( 2 \right)_k$ as:
\begin{multline}
  \frac{2}{k} \di s^2 = \tr \left( \Omega^\dag{} \Omega \right) = - \frac{1 }{2
    f^2}\left( \bar z^2 \di z \otimes \di z + z^2 \di \bar z \otimes \di \bar z
    - 2 \left( 2 + \abs{z}^2 \right) \di z \otimes \di \bar z \right) +
  \\+\frac{\imath }{f} \left( z \di \bar z - \bar z \di z \right) \otimes \di \psi
  + \frac{1}{2} \di \psi \otimes \di \psi
\end{multline}
The left-moving current contains a contribution from the free fermions
realizing an $SU(2)_2$ algebra, so that the theory possesses (local)
$N=(1,0)$ superconformal symmetry. 

The marginal deformation is obtained by switching on a magnetic field in the
$SU(2)$, introducing the following
$(1,0)$-superconformal-symmetry-compatible marginal operator:
\begin{equation}
  \delta S = \frac{\sqrt{k k_g}\textsc{h}}{2\pi} (J^3 + \lambda^+ \lambda^-) \bar{J}
\end{equation}
where we have picked one particular current $\bar{J}$ from the gauge sector,
generating a $U(1)$ at level $k_g$. For instance, we can choose the level
two current: $\bar J = i \tilde{\chi}^1 \tilde{\chi}^2$.
As a result the solutions to the deformed $\sigma $-model~\eqref{eq:def-metric},
\eqref{eq:def-F-field} and \eqref{eq:def-H-field} read:
\begin{align}
  \frac{1}{k} \di s^2 &= \frac{\di z \otimes \di \bar z}{\left( 1 +
      \abs{z}^2\right)^2 } + \left( 1 - 2 \textsc{h}^2 \right) \frac{1}{f^2}\left(
    \imath z \di \bar z - \imath \bar z \di z + f \di \psi \right) \otimes \left( \imath z
    \di \bar z - \imath \bar z \di z +
    f \di \psi \right) \label{eq:deformed-su2-metric}\\
  \di B &= \frac{\imath k}{2} \frac{1}{\left( 1+ \abs{z}^2\right)^2} \di z \land
  \di \bar z \land \di \psi \\
  A &= \sqrt{\frac{k}{2 k_g}}\textsc{h} \left( - \frac{\imath }{f}\left( \bar z \di z - z \di
    \bar z \right)+ \di \psi\right)
\end{align}
It can be useful to write explicitly the volume form on the manifold and the
Ricci scalar:
\begin{align}
  \sqrt{\det g} \ \di z \land \di \bar z \land \di \psi &= \frac{k}{2} \frac{\sqrt{k
      \left( 1 - 2 \textsc{h}^2 \right)}}{\left( 1+ \abs{z}^2 \right)^2} \ \di z \land
  \di \bar z \land \di  \psi \\
  R &= \frac{6 + 4 \textsc{h}^2}{k}
\end{align}

It is quite clear that at $\textsc{h} = \textsc{h}_{\text{max}}= 1/\sqrt{2}$ something happens
as it was already remarked in \cite{Kiritsis:1995iu}. In general the
three-sphere $SU \left( 2\right)$ can be seen a non-trivial fibration of $U
\left( 1\right) \sim S^1 $ as fiber and $SU \left( 2\right)/ U\left( 1\right)
\sim S^2$ as base space: the parameterization in
(\ref{eq:deformed-su2-metric}) makes it clear that the effect of the
deformation consists in changing the radius of the fiber that naively seems
to vanish at $\textsc{h}_{\max}$. But as we already know the story is a bit
different: reparameterising as in Eq.~\eqref{rescale}:
\begin{equation}
  \psi \to  \frac{\hat \psi }{\sqrt{1-2\textsc{h}^2}}
\end{equation}
one is free to take the $\textsc{h} \to 1/ \sqrt{2}$ limit where the background fields
assume the following expressions:
\begin{align}
  \frac{1}{k} \di s^2 &\xrightarrow[\textsc{h} \to \nicefrac{1}{\sqrt{2}}]{} \frac{ \di
    z \otimes \di \bar z}{\left( 1 + \abs{z}^2\right)^2 } + \di \hat \psi \otimes
  \di \hat \psi \\ 
  F &\xrightarrow[\textsc{h} \to \nicefrac{1}{\sqrt{2}}]{} \sqrt{\frac{k}{4 k_g}}\frac{
    \imath \di z \land \di \bar z }{\left( 1 +
      \abs{z}^2\right)^2} \\
  H &\xrightarrow[\textsc{h} \to \nicefrac{1}{\sqrt{2}}]{} 0
\end{align}

Now we can justify our choice of coordinates: the $\left( z, \bar
  z \right)$ part of the metric that decouples from the $\psi $ part is
nothing else than the Kähler metric for the manifold $\mathbb{CP}^1$ (which
is isomorphic to $SU \left( 2\right)/U\left( 1\right)$). In this terms the
field strength $F$ is proportional to the K\"ahler two-form:
\begin{equation}
  F = \imath \, \sqrt{\frac{k}{k_g}} g_{z \bar z  } \ 
  \di z \land \di \bar z 
\end{equation}
This begs for a remark. It is simple to show that cosets of the form $G/H$
where $H$ is the maximal torus of $G$ can always be endowed with a Kähler
structure. The natural hope is then for this structure to pop up out of our
deformations, thus automatically assuring the $N=2$ world-sheet
supersymmetry of the model. Actually this is not the case. The Kähler
structure is just one of the possible left-invariant metrics that can be
defined on a non-symmetric coset (see App.~\ref{sec:coset-space-geometry})
and the obvious generalization of the deformation considered above leads to
$\setC$-structures that are not Kähler. From this point of view this first
example is an exception because $SU \left( 2 \right) / U \left( 1 \right)$
is a symmetric coset since $U \left( 1 \right)$ is not only the maximal
torus in $SU \left( 2 \right)$ but also the maximal subgroup. It is
nonetheless possible to define exact an \textsc{cft} on flag spaces but this
will require a slightly different construction, that we will introduce in
Sec.~\ref{sec:gauging}.

We conclude this section observing that the flux of the gauge field on the
two-sphere is given by:
\begin{equation}
  \mathcal{Q} = \int_{S^2} F = \sqrt{\frac{k}{k_g}} \int d\Omega_2 = 
  \sqrt{\frac{k}{k_g}} 4 \pi
\label{flux}
\end{equation}
However one can argue on general grounds that this flux has to be quantized, 
e.g. because the two-sphere appears as a factor of the magnetic monopole 
solution in string theory~\cite{Kutasov:1998zh}.
This quantization of the magnetic charge is only compatible with
levels of the affine $SU(2)$ algebra satisfying the condition:
\begin{equation}
  \frac{k}{k_g} =p^2 \ , \ \ p \in \zi.
  \label{chargeq}
\end{equation}

\newcommand{\jj}[2]{\mathcal{J}^{#1} \land \mathcal{J}^{#2} }
\setcounter{footnote}{0}
%

\subsection{The SU(3)/U(1) flag space}
\label{sec:su3-u1-u1}

Let us now consider the next example in terms of coset dimensions, $SU
\left( 3 \right) / U \left( 1 \right)^2$. As a possible application for this
construction we may think to associate this manifold to a four-dimensional
$\left( 1,0 \right)$ superconformal field theory $\mathcal{M}$ so to
compactify a critical string theory since $\dim \left[SU \left( 3 \right)/ U
  \left( 1 \right)^2 \right]= 8 - 2 =6$. Our construction gives rise to a
whole family of \textsc{cft}'s depending on two parameters (since $\rank
\left[ SU \left( 3 \right) \right] = 2$) but as before we are mainly
interested to the point of maximal deformation, where the $U \left( 1
\right)^2 $ torus decouples and we obtain an exact theory on the $SU \left(
  3 \right) / U \left( 1 \right)^2 $ coset.  Before giving the explicit
expressions for the objects in our construction it is hence useful to
remember some properties of this manifold. The first consideration to be
made is the fact that $SU \left( 3 \right) / U \left( 1 \right)^2 $ is an
asymmetric coset in the mathematical sense defined in
App.~\ref{sec:coset-space-geometry} (as we show below). This allows for the
existence of more than one left-invariant Riemann metric. In particular, in
this case, if we just consider structures with constant Ricci scalar, we
find, together with the restriction of the Cartan-Killing metric on $SU
\left( 3 \right)$, the K\"ahler metric of the flag space $F^3$. The
construction we present in this section will lead to the first one of these
two metrics. This is known to admit a nearly-K\"ahler structure and has
already appeared in the superstring literature as a basis for a cone of
$G_2$ holonomy \cite{Atiyah:2001qf}.

A suitable parametrisation for the $SU \left( 3 \right)$ group is obtained
via the Gauss decomposition described in App.~\ref{sec:suleft-3right}. In
these terms the general group element is written as:
\begin{equation}
  g \left( z_1, z_2, z_3, \psi_1, \psi_2 \right) = \begin{pmatrix}
    \frac{e^{\imath \psi_1/2}}{\sqrt{f_1}} & - \frac{\bar z_1 + z_2 \bar
      z_3}{\sqrt{f1 f2}} e^{\imath \left( \psi_1 - \psi_2 \right)/2} & - \frac{\bar
    z_3 - \bar z_1 \bar z_2}{\sqrt{f_2}} e^{-\imath \psi_2 /2} \\
  \frac{z_1 e^{\imath \psi_1/2}}{\sqrt{f_1}} & - \frac{1+\abs{z_3}^2 - z_1 z_2
  \bar z_3 }{\sqrt{f1 f2}} e^{\imath \left( \psi_1 - \psi_2 \right)/2} & - \frac{
    \bar z_2}{\sqrt{f_2}} e^{-\imath \psi_2 /2} \\
  \frac{z_3 e^{\imath \psi_1/2}}{\sqrt{f_1}} & - \frac{z_2 - \bar z_1 z_3 +
    z_2 \abs{z_1}^2}{\sqrt{f1 f2}} e^{\imath \left( \psi_1 - \psi_2 \right)/2} &
  \frac{1}{\sqrt{f_2}} e^{-\imath \psi_2 /2} \\ 
\end{pmatrix}
\end{equation}
where $z_i $ are three complex parameters, $\psi_j$ are two real parameters
and $f_1 = 1 + \abs{z_1}^2 + \abs{z_3}^2$, $f_2 = 1 + \abs{z_2}^2 +
\abs{z_3 - z_1 z_2}^2$. As for the group, we need also an explicit
parameterisation for the $\mathfrak{su} \left( 3 \right)$ algebra, such as
the one provided by the Gell-Mann matrices in
Eq.~\eqref{eq:Gell-Mann-matrices}. It is a well known result that if a Lie
algebra is semi-simple (or, equivalently, if its Killing form is
negative-definite) then all Cartan subalgebras are conjugated by some inner
automorphism\footnote{This is the reason why the study of non-semi-simple
  Lie algebra deformation constitutes a richer subject. In example the $SL
  \left( 2, \setR \right)$ group admits for 3 different deformations, leading
  to 3 different families of exact \textsc{cft}'s with different physics
  properties. On the other hand the 3 possible deformations in $SU \left( 3
  \right)$ are equivalent.}. This leaves us the possibility of choosing any
couple of commuting generators, knowing that the final result won't be
influenced by such a choice. In particular, then, we can pick the subalgebra
generated by $\mathfrak{k} = \braket{\lambda_3, \lambda_8}$.\footnote{In this explicit
  parameterisation it is straightforward to show that the coset we're
  considering is not symmetric.  It suffices to pick two generators, say
  $\lambda_2 $ and $\lambda_4$, and remark that their commutator $\left[ \lambda_2, \lambda_4
  \right] = - 1/ \sqrt{2} \lambda_6$ doesn't live in the Cartan subalgebra.}

We can now specialize the general expressions given in
Sec.~\ref{sec:geom-constr}. The holomorphic currents
\eqref{eq:holo-currents} of the bosonic $SU \left( 3 \right)_k$
corresponding to the two operators in the Cartan are:
\begin{align}
  \mathcal{J}^3 = - \tr \left( \lambda_3 g\left( z_\mu , \psi_a \right)^{-1} \di
    g\left( z_\mu , \psi_a \right) \right) && \mathcal{J}^8 = - \tr \left(
    \lambda_8 g\left( z_\mu, \psi_a \right)^{-1} \di g\left( z_\mu , \psi_a \right)
  \right)
\end{align}
that in these coordinates read:
\begin{small}
  \begin{multline}
    \mathcal{J}^3 = - \frac{\imath}{\sqrt{2}} \left\{ \left( \frac{\bar
          z_1}{f_1}+ \frac{ z_2 \left( - \bar z_1 \bar z_2 + \bar z_3
          \right)}{2 f_2} \right)\di z_1 - \frac{ \bar z_2 \left( 1+
          \abs{z_1}^2\right)- z_1 \bar z_3 }{2 f_2} \di z_2 + \left(
        \frac{\bar z_3}{f_1} + \frac{ \bar z_1 \bar z_2 - \bar z_3 }{2
          f_2} \right) \di z_3 \right\} \\+ \text{c.c.} + \frac{\di
      \psi_1}{\sqrt{2}}- \frac{\di \psi_2 }{2 \sqrt{2}}
  \end{multline}
  \begin{equation}
    \mathcal{J}^8 = - \imath \sqrt{\frac{3}{2}} \left\{ \frac{\bar z_1 \bar z_2 -
        \bar z_3 }{2f_2} z_2 \di z_1 + \frac{ \bar z_2 + \abs{z_1}^2
        \bar z_2 - z_1 \bar z_3 }{2f_2} \di z_2 + \frac{ -\bar z_1 \bar
        z_2 + \bar z_3}{2f_2}\di z_3 \right\} + \text{c.c.}+
    \frac{1}{2} \sqrt{\frac{3}{2}} \di \psi_2
  \end{equation}
\end{small}
they appear in the expression of the exactly marginal operator
\eqref{margdef} that we can add to the $SU \left( 3 \right)$ \textsc{wzw}
model action is:
\begin{multline}
    V = \frac{\sqrt{k k_g}}{2 \pi } \textsc{h} \int \di z^2 \: \textsc{h}_3 \left( J^3 - \frac{\imath
      }{\sqrt{2} k}\left( 2 : \psi_2 \psi_1 : + : \psi_5 \psi_4 : + :\psi_7 \psi_6 :
      \right) \right) \bar J^3 +\\+ \textsc{h}_8 \left( J^8 - \frac{\imath }{k}
      \sqrt{\frac{3}{2}} \left( :\psi_5 \psi_4: + :\psi_7 \psi_6:\right)\right) \bar
    J^8
\end{multline}
where $\psi^i $ are the bosonic current superpartners and $\bar J$ are two
currents from the gauge sector both generating a $U \left( 1 \right)_{k_g}$.

Since $\rank \left[ SU \left( 3 \right) \right] = 2 $ we have a
bidimensional family of deformations parameterised by the two moduli $\textsc{h}_3 $
and $\textsc{h}_8$. The back-reaction on the metric is given by:
\begin{equation}
  \di s^2 = g_{\alpha \bar \beta } \di z^\alpha \otimes \di \bar z^\beta + \left( 1 - 2 \textsc{h}_3^2
  \right)  \mathcal{J}^3 \otimes \mathcal{J}^3  + \left( 1 - 2 \textsc{h}_8^2
  \right)  \mathcal{J}^8 \otimes  \mathcal{J}^8 
\end{equation}
where $g_{\alpha \bar \beta }$ is the restriction of the $SU \left( 3 \right)$
metric on $SU \left( 3 \right)/U\left( 1 \right)^2$. It is worth to remark
that for any value of the deformation parameters $\textsc{h}_3 $ and $\textsc{h}_8$ the
deformed metric is Einstein with constant Ricci scalar.

With a procedure that has by now become familiar we introduce the following
reparametrisation:
\begin{align}
  \psi_1 = \frac{\hat \psi_1}{\sqrt{1-2\textsc{h}^2}} && \psi_2 = \frac{\hat \psi_2}{\sqrt{1-2\textsc{h}^2}}
\end{align}
and take the $\textsc{h}_3\to1/ \sqrt{2}$, $\textsc{h}_8\to1/ \sqrt{2}$ limit. The resulting metric
is:
\begin{equation}
   \di s^2 =  g_{\alpha \bar \beta } \di z^\alpha \otimes \di \bar z^\beta + \frac{\di \hat \psi_1 \otimes \di
     \hat \psi_1 - \di \hat \psi_1 \otimes \di \hat \psi_2 + \di \hat \psi_2 \otimes \di \hat \psi_2 }{2}
\end{equation}
that is the metric of the tangent space to the manifold $SU \left( 3
\right)/U\left( 1 \right)^2 \times U\left( 1\right) \times U \left( 1 \right)$. As
shown in App.~\ref{sec:coset-space-geometry} the coset metric hence obtained
has a $\setC$-structure, is Einstein and has constant Ricci scalar $R=15/k$.
The other background fields at the boundary of the moduli space
read:
\begin{gather}
  F = \di \mathcal{J}^3 + \di \mathcal{J}^8  \\
  H_{\left[ 3 \right]} =- 3 \sqrt{2} \left\{ \mathcal{J}^1 \land \left( \jj{4}{5} -
      \jj{6}{7}\right) + \sqrt{3} \mathcal{J}^2 \land \left( \jj{4}{5}
    + \jj{6}{7} \right) \right\}
\end{gather}

  If we consider the supersymmetry properties along the deformation line we
  can remark the presence of an interesting phenomenon. The initial $SU
  \left( 3 \right)$ model has $N=2$ but this symmetry is naively
  broken to $N=1$ by the deformation. This is true for any value
  of the deformation parameter but for the boundary point $\textsc{h}_3^2 = \textsc{h}_8^2 =
  \nicefrac{1}{2}$ where the $N = 2 $ supersymmetry is restored.
  Following \cite{Gates:1984nk,Kazama:1989qp,Kazama:1989uz} one can see that
  a $G/T$ coset admits $N=2$ supersymmetry if it possesses a
  complex structure and the corresponding algebra can be decomposed as
  $\mathfrak{j} = \mathfrak{j}_+ \oplus \mathfrak{j}_-$ such as
  $\comm{\mathfrak{j}_+,\mathfrak{j}_+}=\mathfrak{j}_+$ and
  $\comm{\mathfrak{j}_-,\mathfrak{j}_-}=\mathfrak{j}_-$. Explicitly, this
  latter condition is equivalent (in complex notation) to $f_{ijk} = f_{\bar
    i \bar j \bar k} = f_{a i j} = f_{a \bar i \bar j} = 0$. These are
  easily satisfied by the $SU \left( 3 \right)/ U \left( 1 \right)^2 $ coset
  (and actually by any $G/T $ coset) since the commutator of two positive
  (negative) roots can only be proportional to the positive (negative) root
  obtained as the sum of the two or vanish, as shown in
  Eq.~\eqref{eq:CartanWeyl}. Having $N=2$ supersymmetry is
  equivalent to asking for the presence of two complex structures. The first
  one is trivially given by considering positive roots as holomorphic and
  negative roots as anti-holomorphic, the other one by interchanging the role in one
  out of the three positive/negative couples (the same flip on two couples
  would give again the same structure and on all the three just takes back
  to the first structure). The metric is hermitian with respect to both
  structures since it is $SU \left( 3\right)$ invariant. It is worth to
  remark that such background is different from the ones described in
  \cite{Kazama:1989uz} because it is not K\"ahler and can't be decomposed in
  terms of Hermitian symmetric spaces.

\newcommand{\FF}[2]{{F^{#1}_{\phantom{#1}#2}}}
\setcounter{footnote}{0}

\section{Gauging}
\label{sec:gauging}

In this section we want to give an alternative construction for our deformed
models, this time explicitly based on an asymmetric \textsc{wzw} gauging.
The existence of such a construction is not surprising at all since our
deformations can be seen as a generalization of the ones considered in
\cite{Giveon:1994ph}. In these terms, just like $J \bar J$ (symmetric)
deformations lead to gauged \textsc{wzw} models, our asymmetric construction
leads to asymmetrically gauged \textsc{wzw} models, which were studied in
\cite{Quella:2002fk}.

First of all we will give the explicit construction for the most simple
case, the $SU \left( 2 \right)$ model, then introduce a more covariant
formalism which will be simpler to generalize to higher groups, in
particular for the $SU \left( 3 \right)$ case, whose gauging will lead, this
time, to two different exact models corresponding to the two possible
Einstein complex structures admitted by the $SU \left( 3 \right) / U \left(
  1 \right)^2 $ manifold.

To simplify the formalism we will discuss gauging of bosonic \textsc{cft}s, 
and the currents of the gauge sector of the heterotic string are replaced 
by compact $U(1)$ free bosons. It is obvious that all the results 
are easily translated into heterotic string constructions.

\subsection{The SU(2)/U(1) asymmetric gauging}
\label{sec:su-left-2}

In this section we want to show how the $S^2 $ background described in
\cite{Israel:2004vv} can be directly obtained via an asymmetric gauging of
the $SU \left( 2 \right) × U \left( 1 \right)$ \textsc{wzw} model (a similar
construction was first obtained in \cite{Johnson:1995kv}).

Consider the \textsc{wzw} model for the group manifold $SU \left( 2 \right)_k × U
\left( 1 \right)_{k^\prime}$. A parametrisation for the general element of this
group which is nicely suited for our purposes is obtained as follows:
\begin{equation}
  g = 
  \begin{pmatrix}
    z_1 & z_2 & 0 \\
    - \bar z_2 & \bar z_1 & 0 \\
    0 & 0 & z_3
  \end{pmatrix} = \left( \begin{tabular}{c|c}
        $g_2 $& $ 0$\\ \hline
        $ 0 $& $g_1$
      \end{tabular}\right) \in SU \left( 2 \right) × U \left( 1 \right) 
\end{equation}
where $g_1 $ and $g_2 $ correspond to the $SU \left( 2 \right)$
and $U \left( 1 \right)$ parts respectively and $\left( z_1, z_2, z_3 \right)$ satisfy:
\begin{equation}
  SU \left( 2 \right) × U \left( 1 \right) = \set{ \left( w_1, w_2,
      w_3 \right) | \abs{w_1}^2 + \abs{w_2}^2 = 1, \abs{w_3}^2 = 1} \subset \setC^3  
\end{equation}
A possible choice of coordinates for the corresponding group manifold is
given by the Euler angles:
\begin{multline}
\label{eq:su2u1coords}
  SU \left( 2 \right) × U \left( 1 \right) \\= \Set{\left( z_1, z_2, z_3
    \right) 
= \left( \cos \frac{\beta }{2} e^{ \imath \left( \gamma + \alpha \right)/2 },
      \sin \frac{\beta }{2} e^{\imath \left( \gamma - \alpha \right)/2}, e^{\imath \varphi}\right)|
    0 \leq \beta  \leq \pi , 0 \leq \alpha , \beta , \varphi \leq 2 \pi }
\end{multline}

In order to obtain the coset construction leading to the $S^2 $
background we define two $U \left( 1 \right) \to SU \left( 2 \right) ×
U\left( 1 \right)$ embeddings as follows:
\begin{align}
\label{eq:su2-embeddings}
  \begin{split}
    \epsilon_L :U \left( 1 \right) &\to SU \left( 2 \right) × U \left( 1 \right) \\
    e^{\imath \tau} &\mapsto \left( e^{\imath \tau}, 0, 1\right)
  \end{split}
  \begin{split}
    \epsilon_R :U \left( 1 \right) &\to SU \left( 2 \right) × U \left( 1 \right) \\
    e^{\imath \tau} &\mapsto \left( 1, 0, e^{\imath \tau}\right)
  \end{split}
\end{align}
so that in terms of the $z$ variables the action of these embeddings boils
down to:
\begin{align}
  g &\mapsto \epsilon_L \left( e^{\imath \tau } \right) g \epsilon_R \left( e^{\imath \tau }\right)^{-1} \\
  \left( w_1, w_2, w_3 \right) & \mapsto \left( e^{\imath \tau } w_1, 
    e^{\imath \tau } w_2, e^{-\imath \tau } w_3 \right)
\end{align}
This means that we are free to choose a gauge where $w_2$ is real or, in
Euler coordinates, where $\gamma = \alpha $, the other angular variables
just being redefined.  To find the background fields corresponding to this
gauge choice one should simply write down the Lagrangian where the
symmetries corresponding to the two embeddings in \eqref{eq:su2-embeddings}
are promoted to local symmetries, integrate the gauge fields out and then
apply a Kaluza-Klein reduction, much in the same spirit as in
\cite{Israel:2004vv}.

The starting point is the \textsc{wzw} model, written as:
\begin{equation}
  S_{\textsc{wzw}} \left( g \right) = \frac{k}{4 \pi } \int \di z^2 \: \Tr
  \left( g_2^{-1} \d g_2 g_2^{-1} \db g_2 \right) + \frac{k^\prime }{4 \pi } \int \di z^2 \: \Tr
  \left( g_1^{-1} \d g_1 g_1^{-1} \db g_1 \right)
\end{equation}
Its gauge-invariant generalization is given by:
\begin{multline}
  S = S_{\textsc{wzw}}\\ + \frac{1}{2 \pi } \int \di^2 z \left[ k \bar A \Tr \left(
    t_L \partial g g^{-1}\right) + k^\prime A \Tr \left( t_R g^{-1} \bar \partial g\right) +
  \sqrt{k k^\prime }A
  \bar A \left( -2 + \Tr \left( t_L \, g \, t_R \, g^{-1}\right)\right) \right]
\end{multline}
where $A $ and $\bar A $ are the components of the gauge field, and 
$t_L$ and $t_R$ are the Lie algebra generators corresponding to 
the embeddings in
\eqref{eq:su2-embeddings}, \emph{i.e.}
\begin{align}
  t_L = \imath \left( \begin{tabular}{c|c}
      $\sigma_3 $& $  0 $\\ \hline
      $ 0 $& $0$
    \end{tabular}\right), &&
  t_R = \imath \left( \begin{tabular}{c|c}
      $0$ & $0$ \\ \hline
      $0$ & $p$
    \end{tabular}\right),
\end{align}
$\sigma_3 $ being the usual Pauli matrix. For such an asymmetric coset to 
be anomaly free, one has the following constraint on the embeddings:
\begin{equation}
k \Tr (t_L)^2 = k' \Tr (t_R)^2 \implies 
k = k' p^2 \ , \ \ \text{with} \ p \in \mathbb{N}.
\label{anomalys2}
\end{equation}
If we pass to Euler coordinates it is
simple to give an explicit expression for the action:
\begin{multline}
  S \left( \alpha, \beta, \gamma, \varphi \right) = \frac{1}{2 \pi } \int \di^2 z \: \frac{k}{4} \left( \partial \alpha \bar \partial \alpha  +
    \partial \beta \bar \partial \beta  + \partial \gamma \bar \partial \gamma + 2 \cos \beta \partial \alpha \bar \partial \gamma
  \right) + \frac{k^\prime }{2} \partial
    \varphi \bar \partial \varphi + \\+ \imath k \left( \partial \alpha + \cos \beta \partial \gamma \right) \bar A +
  \imath k^\prime \sqrt{2} \bar \partial \varphi  A - 2 \sqrt{k k^\prime } A \bar A 
\end{multline}
This Lagrangian is quadratic in $A, \bar A $ and the quadratic part is
constant so we can integrate these gauge fields out and the resulting
Lagrangian is:
\begin{multline}
\label{eq:gauged-SU2}
  S \left( \alpha, \beta, \gamma, \varphi \right) = \frac{1}{2 \pi } \int \di^2 z \: \frac{k}{4} \left( \partial \alpha \bar \partial \alpha  +
    \partial \beta \bar \partial \beta  + \partial \gamma \bar \partial \gamma + 2 \cos \beta \partial \alpha \bar \partial \gamma \right) + \frac{k^\prime }{2} \partial
    \varphi \bar \partial \varphi + \\+ \frac{\sqrt{2 k k^\prime }}{2} \left( \d \alpha + \cos \beta \d
    \gamma \right) \db \varphi 
\end{multline}
now, since we gauged out the symmetry corresponding to the $U \left( 1
\right)$ embeddings, this action is redundant. This can very simply be seen
by writing the corresponding metric and remarking that it has vanishing
determinant: 
\begin{equation}
  \det g_{\mu \nu } =  
  \begin{vmatrix}
    k/4 \\
    & k/4 &k/4 \cos \beta  & \sqrt{2 k k^\prime }/4\\
    & k/4 \cos \beta & k/4 & \sqrt{2 k k^\prime }/4 \cos \beta \\ 
    &  \sqrt{2 k k^\prime }/4 & \sqrt{2 k k^\prime }/4 \cos \beta  & k^\prime/2 
  \end{vmatrix} = 0
\end{equation}
Of course this is equivalent to say that we have a gauge to fix (as we saw
above) and this can be chosen by imposing $\gamma = \alpha $, which leads to the
following action:
\begin{equation}
  S \left( \alpha, \beta, \varphi \right) = \frac{1}{2 \pi } \int \di^2 z \: \frac{k}{4}
  \left( 2 \left( 1 + \cos \beta \right) \partial \alpha \bar \partial \alpha  +
    \partial \beta \bar \partial \beta  \right) + \frac{k^\prime }{2} \partial
    \varphi \bar \partial \varphi+ \frac{\sqrt{2 k k^\prime }}{2} \left( 1 + \cos \beta
    \right) \d \alpha \db \varphi 
\end{equation}
whence we can read a two dimensional metric by interpreting the $\d \alpha \db
\varphi $ term as a gauge boson and applying the usual Kaluza-Klein reduction. We
thus recover the two-sphere we expect.
\begin{equation}
  \label{eq:S2-line-element}
  \di s^2 = g_{\mu \nu } - G_{\varphi \varphi } A_\mu A_\nu = \frac{k}{4}\left( \di \beta^2 +
  \sin^2 \beta \di \alpha^2 \right)
\end{equation}
supported by a (chromo)magnetic field
\begin{equation}
\label{eq:S2-magnetic-field}
  A = \sqrt{\frac{k} {k^\prime} } \left( 1 + \cos \beta \right) \di \alpha   
\end{equation}

As advertised above we now turn to rewrite the above gauging in a more
covariant form, simpler to generalize. Since we're interested in the
underlying geometry, we'll mainly focus on the metric of the spaces we
obtain at each step and write these metrics in terms of the Maurer-Cartan
currents\footnote{One of the advantages of just working on the metrics is
  given by the fact that in each group one can consistently choose
  holomorphic or anti-holomorphic currents as a basis. In the following we
  will consider the group in the initial \textsc{wzw} model as being
  generated by the holomorphic and the dividing group by the
  anti-holomorphic ones.}. As we've already seen in Eq.~\eqref{eq:carkil},
the metric of the initial group manifold is:
\begin{equation}
  \di s^2 = \frac{k}{2} \sum \mJ_i^2 \otimes \mJ_i^2 + \frac{k^\prime }{2} \mI \otimes \mI  
\end{equation}
where $\set{\mJ_1, \mJ_2, \mJ_3 }$ are the currents of the $SU \left( 2 \right)$
part and $\mI$ the $U \left( 1 \right)$ generator. The effect of the
asymmetric gauging amounts - at this level - to adding what we can see as an
interaction term between the two groups. This changes the metric to:
\begin{equation}
  \di s^2 = \frac{k}{2} \sum \mJ_i^2 \otimes \mJ_i^2 + \frac{k^\prime }{2} \mI \otimes \mI  +
  \sqrt{k k^\prime } \mJ_3 \otimes \mI    
\end{equation}
Of course if we choose $\braket{\mJ_1, \mJ_2, \mJ_3, \mI}$ as a basis we can
rewrite the metric in matrix form:
\begin{equation}
  g = \frac{1}{2} 
  \begin{pmatrix}
    k \\
    & k \\
    & & k & \sqrt{k k^\prime }\\
    & & \sqrt{k k^\prime } & k^\prime 
  \end{pmatrix}
\end{equation}
where we can see that the gauging of the axial symmetry corresponds to the
fact that the sub-matrix relative to the $\set{\mJ_3, \mI }$ generators is
singular:
\begin{equation}
  \begin{vmatrix}
    k & \sqrt{k k^\prime } \\
    \sqrt{k k^\prime } & k^\prime 
  \end{vmatrix} = 0
\end{equation}
explicitly this correspond to:
\begin{equation}
  k \mJ_3 \otimes \mJ_3 + \sqrt{k k^\prime } \mJ_3 \otimes \mI + \sqrt{k k^\prime } \mJ_3 \otimes \mI + k^\prime \mI
  \otimes \mI = \left( k + k^\prime \right) \hat \mJ \otimes \hat \mJ
\end{equation}
where
\begin{equation}
  \hat \mJ = \frac{\sqrt{k} \mJ_3 + \sqrt{k^\prime } \mI}{\sqrt{k + k^\prime }}  
\end{equation}
is a normalized current. In matrix term this corresponds to projecting the
interaction sub-matrix on its non-vanishing normalized eigenvector:
\begin{equation}
  \begin{pmatrix}
    \sqrt{\frac{k}{k+k^\prime }} & \sqrt{\frac{k}{k+k^\prime }}
  \end{pmatrix}
  \begin{pmatrix}
     k & \sqrt{k k^\prime } \\
    \sqrt{k k^\prime } & k^\prime
  \end{pmatrix}
  \begin{pmatrix}
    \sqrt{\frac{k}{k+k^\prime }} \\ \sqrt{\frac{k}{k+k^\prime }}
  \end{pmatrix} = k + k^\prime 
\end{equation}
and the resulting metric in the $\braket{\mJ_1, \mJ_2, \hat \mJ }$ basis is:
\begin{equation}
\label{eq:asy-gaug-su2}
  \begin{pmatrix}
    k \\
    & k \\
    & & k+k^\prime 
  \end{pmatrix}
\end{equation}
This manifold $\mathcal{M}$ (whose metric appears in the action
\eqref{eq:KK-action}) corresponds to a $S^1 $ fibration (the fiber being
generated by $\hat \mJ$) over a $S^2 $ base (generated by $\braket{\mJ_1, \mJ_2
}$).
\begin{equation}
  \begin{CD}
    S^1 @>>> \mathcal{M} \\
    @.      @VVV\\
    {} @. S^2
  \end{CD}
\end{equation}

It should now appear obvious how to generalize this construction so to
include all the points in the moduli space joining the unperturbed and
gauged model. The decoupling of the $U \left( 1 \right)$ symmetry (that has
been ``gauged away'') is obtained because the back-reaction of the gauge
field Eq.~\eqref{eq:gauged-SU2} is such that the interaction sub-matrix is
precisely singular. On the other hand we can introduce a parameter that
interpolates between the unperturbed and the gauged models so that the
interaction matrix now has two non-null eigenvalues, one of which will
vanish at the decoupling point. 

In practice this is done by adding to the  
the asymmetrically gauged \textsc{wzw} model an auxilliary 
$U(1)$ free boson $Y$ at radius $R= (k k^\prime )^{\nicefrac{1}{4}}
(\nicefrac{1}{\sqrt{2}\textsc{h}-1})^{\nicefrac{1}{2}}$. 
This $U(1)$ is coupled symmetrically to the gauge fields such that 
the anomaly cancelation condition is still given by~(\ref{anomalys2}). 
In particular if we choose the gauge $Y=0$, the metric reads:
\begin{equation}
  \begin{pmatrix}
    k & \sqrt{2} \textsc{h} \sqrt{k k^\prime } \\
    \sqrt{2} \textsc{h} \sqrt{k k^\prime } & k^\prime 
  \end{pmatrix}
\end{equation}
which is exactly the model studied above. For a generic value of $\textsc{h}^2
$ the two eigenvalues are given by:
\begin{equation}
  \label{eq:Interact-Eigenv}
  \lambda_{1\atop2} \left( k, k^{\prime }, \textsc{h} \right) = 
\frac{k + k^\prime \mp  \sqrt{k^2 + {k^\prime
      }^2 + 2 \left( 4 \textsc{h}^2 - 1 \right)k k^\prime }}{2}
\end{equation}
so we can diagonalize the metric in the $\braket{\mJ_1, \mJ_2, \hat \mJ, \hat
  {\hat \mJ}}$ basis ($\hat \mJ $ and $\hat {\hat \mJ}$ being the two
eigenvectors) and finally obtain:
\begin{equation}
  g = 
  \begin{pmatrix}
    k \\
    & k \\
    & & \lambda_1\left( k, k^{\prime }, \textsc{h} \right) \\
    & & & \lambda_2\left( k, k^{\prime }, \textsc{h} \right)
  \end{pmatrix}
\end{equation}
Of course, in the $\textsc{h}^2 \to 0$ limit we get the initial \textsc{wzw} model and
in the $\textsc{h}^2 \to 1/2 $ limit we recover the 
asymmetrically gauged model
Eq.~\eqref{eq:asy-gaug-su2}.

It is important to remark that the construction above can be directly
generalized to higher groups with non-abelian subgroups, at least for the
asymmetric coset part. This is what we will do in the next section.

\boldmath
\subsection{SU (3) /U(1)$^2$}
\label{sec:su-3-}
\unboldmath

To study the $SU \left( 3 \right)$ case we will use the ``current''
approach, since a direct computation in coordinates would be impractical. As
one could expect, the study of $SU \left( 3 \right)$ deformation is quite
richer because of the presence of an  embedded $SU \left( 2
\right)$ group that can be gauged. Basically this means that we can choose
two different deformation patterns that will lead to the two possible
Einstein structures that can be defined on the $SU \left( 3 \right) / U
\left( 1 \right)^2$ manifold (see App.~\ref{sec:coset-space-geometry}).

\subsubsection{Direct gauging.}

The first possible choice consists in the obvious generalization of the $SU
\left( 2 \right)/ U \left( 1 \right)$ construction above, \emph{ie} simply
gauging the $U \left( 1 \right)^2 $ Cartan torus. Consider the initial $SU
\left( 3 \right)_k × U \left( 1 \right)_{k^\prime } × U \left( 1 \right)_{k^{\prime \prime
  }}$ model. In the $\braket{\mJ_1,\ldots, \mJ_8, \mI_1, \mI_2}$ base
($\set{\mJ_i}$ being the $SU \left( 3 \right)$ generators and $\set{\mI_k}$
the 2 $U\left( 1 \right)$'s), the initial metric is written as:
\begin{equation}
  g = \left( \begin{tabular}{c|c}
        $k \uni{8} $& $ 0$\\ \hline
        $ 0 $& $
        \begin{matrix}
          k^\prime \\
          & k^{\prime \prime }
        \end{matrix}$
      \end{tabular}\right)
\end{equation}
the natural choice for the Cartan torus is given by the usual
$\braket{\mJ_3, \mJ_8 }$ generators, so we can proceed as before and write
the deformed metric as:
\begin{equation}
  g = 
  \begin{pmatrix}
    k \uni{2} \\
    & \lambda_1 \left( k, k^\prime, \textsc{h}_3 \right) \\
    & & k \uni{4} \\
    & & & \lambda_1 \left( k, k^{\prime\prime }, \textsc{h}_8\right) \\ 
    & & & & \lambda_2 \left( k, k^\prime, \textsc{h}_3 \right) \\
    & & & & & \lambda_2 \left( k, k^{\prime \prime}, \textsc{h}_8 \right) \\    
  \end{pmatrix}
\end{equation}
where $\textsc{h}_3 $ and $\textsc{h}_8 $ are the deformation parameters and $\lambda_1$ and $\lambda_2
$ are the eigenvalues for the interaction matrices, given in
Eq.~\ref{eq:Interact-Eigenv}. In particular, then, in the $\textsc{h}_3^2 \to 1/2 $,
$\textsc{h}_8^2 \to 1/2$ limit two eigenvalues vanish, the corresponding directions
decouple and we're left with the following (asymmetrically gauged) model:
\begin{equation}
  g = \left( \begin{tabular}{c|c}
      $k \uni{6} $ \\ \hline
      & $ \begin{matrix}
        k + k^\prime \\
        & k + k^{\prime \prime }
      \end{matrix}$
    \end{tabular}\right)
\end{equation}
in the $\braket{\mJ_1, \mJ_2, \mJ_4, \mJ_5, \mJ_6, \mJ_7, \sqrt{k^\prime } \mI_1
  + \sqrt{k} \mJ_3, \sqrt{k^{\prime \prime }} \mI_2 + \sqrt{k} \mJ_8 }$ basis that can
be seen as a $U \left( 1 \right)^2$ fibration over a $SU \left( 3 \right) /
U \left( 1 \right)^2 $ base with metric $\diag \left( 1,1,1,1,1,1 \right)$
(in the notation of App~\ref{sec:coset-space-geometry}). This is precisely
the same result we obtained in Sec.~\ref{sec:su3-u1-u1} when we read the
fibration as a gauge field living on the base.
\begin{equation}
  \begin{CD}
    U \left( 1 \right)^2 @>>> \mathcal{M} \\
    @.      @VVV\\
    {} @. SU \left( 3 \right)/ U \left( 1 \right)^2
  \end{CD}
\end{equation}
As in the previous example all this construction is valid only if the 
asymmetrically gauged \textsc{wzw} model is anomaly-free. 
This will be explained in detail in section~\ref{sec:modular-invariance}.

\boldmath
\subsubsection{The F$_3$ flag space}
\unboldmath

Let us now turn to the other possible choice for the $SU \left( 3 \right)$
gauging, namely the one where we take advantage of the $SU \left( 2 \right)$
embedding. Let us then consider the $SU \left( 3 \right)_{k_3} × SU \left( 2
\right)_{k_2} × U \left( 1 \right)_{k^\prime } × U \left( 1 \right)_{k^{\prime \prime
  }}$ \textsc{wzw} model whose metric is
\begin{equation}
  g =  \left( \begin{tabular}{c|c|cc}
      $k_3 \uni{8}$ & & &\\  \hline
      & $k_2 \uni{3}$ & &\\  \hline
      & & $k^\prime$ & \\
      & &  & $k^{\prime \prime }$      
    \end{tabular}\right)
\end{equation}
in the $\braket{\mJ_1, \ldots, \mJ_8, \mI_1, \mI_2, \mI_3, \mK_1, \mK_2}$ basis, where
$\braket{\mJ_i}$ generate the $SU\left( 3 \right)$, $\braket{\mI_i}$ generate
the $SU \left( 2 \right)$ and $\braket{\mK_i} $ generate the $U \left( 1
\right)^2 $.

The first step in this case consists in an asymmetric gauging mixing the
$\set{\mJ_1, \mJ_2, \mJ_3 }$ and $\set{\mI_1, \mI_2, \mI_3}$ currents respectively. At
the gauging point, a whole 3-sphere decouples and we obtain the following
metric:
\begin{equation}
  g =  \left( \begin{tabular}{c|c|cc}
      $k_3 \uni{5}$ & & &\\  \hline
      & $\left( k_2 + k_3 \right) \uni{3}$ & &\\  \hline
      & & $k^\prime$ & \\
      & &  & $k^{\prime \prime }$      
    \end{tabular}\right)
\end{equation}
where we have to remember that in order to have an admissible embedding $k_2
= k_3 = k$. Our result is again -- not surprisingly -- a $SU \left( 2
\right)$ fibration over a $SU \left( 3 \right) / SU \left( 2 \right)$ base
(times the two $U \left( 1 \right)$'s).
\begin{equation}
  \begin{CD}
    SU \left( 2 \right) @>>> \mathcal{M} \\
    @.      @VVV\\
    {} @. SU \left( 3 \right)/ SU \left( 2 \right)
  \end{CD}
\end{equation}

Of course one could be tempted to give $\mathcal{M}$ the same interpretation as
before, namely a $SU \left( 3 \right) / SU\left( 2 \right)$ space supported
by a chromo-magnetic $SU \left( 2 \right)$ field (or, even better, gauging
an additional $U \left( 1 \right)$, of a $\setC \mathbb{P}^2 $ background
with a $SU \left( 2 \right)× U\left( 1 \right)$ chromo-magnetic field).
Actually this is not the case. The main point is the fact that this $SU
\left( 3 \right) × SU \left( 2 \right)$ model is essentially different
from the previous ones because the $U \left( 1 \right)$ factors were the
result of the bosonisation of the right-moving gauge current which in this
way received a (fake) left-moving partner as in Sec.~\ref{sec:asymm-deform}.
This is not possible in the non-abelian case since one can't obtain a $SU
\left( 2 \right)$ at arbitrary level $k$ out of the fermions of the
theory\footnote{This would be of course be possible if we limited ourselves
to small values of $k$, but in this case the whole geometric
interpretation of the background would be questionable. However for 
Gepner-like string compactifications this class of models is 
relevant.}. In other words,
the $SU\left( 2 \right)$ factor is in this case truly a constituent of the
theory and there is no reason why it should be decoupled or be given a
different interpretation from the $SU \left( 3 \right)$ part.  This is why
the structure obtained by the $SU \left( 2 \right)$ asymmetric gauging is to
be considered a 8-dimensional space admitting a $SU \left( 2 \right) \to SU
\left( 3 \right)/SU \left( 2 \right)$ fibration structure, or, equivalently,
a deformed $SU \left( 3 \right)$ where an embedded $SU \left( 2 \right)$ is
at a level double with respect to the other generators.

On the other hand we are still free to gauge away the two $U \left( 1
\right)$ factors just as before. This time we can choose to couple $K_1$
with the $\mJ_8$ factor that was left untouched in the initial $SU \left( 3
\right)$ and $\mK_2 $ with the $\mJ_3+  \mI_3 $
generator. Again we find a two-parameter family of deformations whose metric
can be written as:
\begin{equation}
  g =  \left( \begin{tabular}{c|c|c|ccc}
      $k \uni{4}$ & & &\\  \hline
      & $\mu_1 $ & &\\ \hline
      & & $2 k \uni{2}$ & \\  \hline
      & & & $\nu_1$ \\
      & & & & $\mu_2$ & \\
      & & & & & $\nu_2$      
    \end{tabular}\right)
\end{equation}
where:
\begin{align}
  \mu &= \lambda \left( k, k^\prime, \textsc{h}^\prime \right) \\
  \nu &= \lambda \left( 2 k, k^{\prime \prime}, \textsc{h}^{\prime \prime } \right)
\end{align}
In particular now we can take the decoupling $\textsc{h}^\prime = \textsc{h}^{\prime \prime } \to 1/2 $
limit where we obtain:
\begin{equation}
  g =  \left( \begin{tabular}{c|c|cc}
      $k \uni{4}$ & &  \\  \hline
      & $2 k \uni{2}$ & \\  \hline
      & & $k + k^\prime $ \\
      & & & $2 k + k^{\prime \prime }$
    \end{tabular}\right)
\end{equation}
this structure is once more a $U \left( 1 \right)^2 \to SU \left( 3
\right)/U\left( 1\right)^2 $ fibration but in this case it is perfectly fine
to separate the space components from the gauge field ones. So we can read
out our final background fields as the Kähler metric on $F_3 $ (see
App~.\ref{sec:coset-space-geometry}) supported by a $U \left( 1 \right)^2 $
(chromo)magnetic field.

To summarize our results we can say that the two Einstein structures that
one can define on $SU \left( 3 \right) / U \left( 1 \right)^2$ are both
exact string theory backgrounds:
\begin{itemize}
\item The first one, obtained as the asymmetric coset 
$\frac{SU \left( 3 \right) × U \left(
      1 \right)^2}{U \left( 1 \right)^2}$ is supported by an \textsc{ns-ns} field strength 
  and a magnetic field;
\item The second, corresponding to the $\frac{SU \left( 3 \right) × SU
    \left( 2 \right) × U \left( 1 \right)^2}{SU \left( 2 \right) ×
    U \left( 1\right)^2}$ asymmetric coset is Kähler and hence supported by the
  (chromo-)magnetic field alone.
\end{itemize}

This Kähler structure has been deeply studied both from the mathematical and
physical points of view. In particular the Kähler form can be written as in
App.~\ref{sec:suleft-3right}:
\begin{equation}
  K \left( \gamma_\mu, \bar \gamma_\mu \right) = \log \left[ 1 + \abs{\gamma_1}^2 + \abs{\gamma_3}^2
  \right] + \log \left[ 1 + \abs{\gamma_2}^2 + \abs{\gamma_3 - \gamma_1 \gamma_2 }^2
  \right] 
\end{equation}
It is immediate to show that this manifold is Einstein and in particular its
Ricci scalar is $R = 12 $. Being Kähler, $F_3 $ is torsionless, that means
in turn that there is no \textsc{ns-ns} form\footnote{To be precise one
  could define a $B$ field but this would have to be closed}. Moreover there
is no dilaton by construction\footnote{The dilaton would basically measure
  the difference between the asymmetric coset volume form and the
  homogeneous space one as it is shown in \cite{Tseytlin:1994my}}. The only
other field that supports the background comes from the $U \left( 1
\right)^2$ fibration. Since the manifold is Kähler it is useful to take
advantage of the complex structure and write our background fields in
complex formalism. In these terms the metric is written as:
\begin{equation}
  g = \frac{k}{2} \left( \mJ^1 \otimes \mJ^{\bar 1} + \mJ^2 \otimes \mJ^{\bar 2} +
    2 \mJ^3 \otimes \mJ^{\bar 3} \right)  
\end{equation}
where $\mJ^i $ and $\bar \mJ^{\bar i}$ are the Maurer-Cartan corresponding
to positive and negative roots respectively and the field strength is given
by:
\begin{equation}
  F^a = \sqrt{\frac{k}{2 k_g }} \F{a}{\mu \bar \rho } C^{\bar \rho \sigma
  } R_{\sigma \bar \nu } \mJ^\mu \land  \mJ^{\bar \nu}
\end{equation}
where $C $ is the following tensor
\begin{equation}
  C = \sum_\alpha \mJ^\alpha \otimes \mJ^{\bar \alpha }
\end{equation}

In Sec.~\ref{sec:flag-equations-motion} we show that the metric and
(chromo)magnetic field solve the first order in $\alpha^\prime $ equations of motion.

\setcounter{footnote}{0}
\section{Exact construction: partition functions}
\label{sec:modular-invariance}

In this section we will compute the one-loop partition functions for the
various asymmetric deformations leading to geometric cosets.  We consider
the part of the partition function of the \textsc{cft} affected by the deformation.We have holomorphic supersymmetric characters and anti-holomorphic bosonic
characters of the affine Lie algebra $\hat{\mathfrak{g}}_k$, 
times some anti-holomorphic fermionic characters
from the gauge sector:
\begin{equation}
  Z \oao{a;[h]}{b;[g]} = 
  \sum_{\Lambda , \bar \Lambda} M^{\Lambda \bar \Lambda} 
  \chi^{\Lambda} (\tau ) \ 
  \left( \frac{\vartheta\oao{a}{b} (\tau )}{\eta (\tau )} \right)^{
    \mathrm{dim} ( \mathfrak g ) /2} 
  \bar{\chi}^{\bar \Lambda} \prod_{\ell} 
\bar{\vartheta} \oao{h_\ell}{g_\ell}
\end{equation}
where $(a,b)$ and $(h_\ell ,g_\ell)$ are the spin structures of the (left and
right) worldsheet fermions.  Useful formulas about characters 
is provided in appendix~\ref{appLie}. Starting from the \textsc{cft}s 
defined by these partition functions we will perform the magnetic deformation 
that has been discussed in the previous sections from the 
geometrical point of view.

\boldmath
\subsection{The SU(3)/U(1)$^2$ flag space CFT}
\unboldmath

The partition function for the asymmetric deformation 
of $SU(2)$ has already been given in
\cite{Israel:2004vv}. We can hence begin with the next non-trivial example
of $SU(3)$. In this case we will compare explicitly two
possible constructions, the Kazama-Suzuki method and the 
direct deformation along the Cartan torus to
show that they give the two inequivalent metrics on the geometric coset.

\subsubsection{The Kazama-Suzuki decomposition of SU(3)}
We would like to decompose our \textsc{wzw} model in terms 
of Kazama-Suzuki (\textsc{ks}) cosets, which are conformal theories 
with extended $N=2$ superconformal 
symmetry~\cite{Kazama:1989qp,Kazama:1989uz}.

The simplest of those models 
are the $N=2$ minimal models that are given by the quotient: 
$\nicefrac{SU(2)_{k-2} \times
  SO(2)_1}{U\left( 1 \right)_{k}}$, 
and their characters come from the branching relation:
\begin{equation}
  \chi^{j}_{k-2} \Xi^{s_2}_{2} = 
  \sum_{m \in \zi_{2k}} \mathcal{C}^{j\, (s_2)}_{m} \frac{\Theta_{m,k}}{\eta}
\end{equation}
For convenience, we write the contribution of the worldsheet 
fermions in terms of  $SO(2n)_1$ characters, see appendix~\ref{appLie}.

Similarly it is possible to construct an $N=2$ coset
\textsc{cft} from $SU(3)$~\cite{Kazama:1989qp,Kazama:1989uz}:\footnote{ 
According to our conventions, the 
weights of a $U\left( 1 \right)$ at level $k$ are
  $\nicefrac{m^2}{4k}$, $m \in \zi_{2k}$.}
\begin{equation}
  \frac{SU\left(3\right)_{k-3} \times SO(4)_1}{SU(2)_{k-2} \times 
    U\left( 1 \right)_{3k}}.
\end{equation}
The characters of this theory are implicitly defined by the branching
relation:
\begin{equation}
\chi^{\Lambda}_{k-3} \, \Xi^{s_4}_{4}  = 
\sum_{2j=0}^{k-2} \sum_{n \in \zi_{6k}} \mathcal{C}^{\Lambda \, (s_4 )}_{j\, n} 
\chi^{j}_{k-2} \, \frac{\Theta_{n,3k}}{\eta}
\end{equation}
Therefore combining the two branching relations, we obtain the decomposition 
of $SU \left( 3 \right)$ in terms of $N=2$ \textsc{ks} models:
\begin{equation}
\chi^{\Lambda}_{k-3} \,  
 \Xi^{s_4}_{4} \Xi^{s_2}_{2}  = 
\sum_{j,m,n}  \mathcal{C}^{\Lambda \, (s_4)}_{j\, n}
\mathcal{C}^{j\, (s_2)}_{m} \ \frac{\Theta_{m,k}}{\eta} \ 
\frac{\Theta_{n,3k}}{\eta}\  
\end{equation}
This decomposition follows the following pattern:
\begin{equation}
SU\left(3\right)_{k-3} \times SO(8)_1 \to 
\frac{SU\left(3\right)_{k-3} \times SO(4)_1}{SU(2)_{k-2} \times 
U\left( 1 \right)_{3k}} \times \frac{SU(2)_{k-2} \times SO(2)_1}{U\left( 1 \right)_{k}}
\times U\left( 1 \right)_{3k} \times U\left( 1 \right)_k \times SO(2)_1
\label{decompA3}
\end{equation}
and we shall perform the deformation on the 
left lattice of $U\left( 1 \right)_{3k} \times U\left( 1 \right)_k$. However 
the deformation will also act on an appropriate 
sub-lattice of the right-moving gauge sector. The last 
$SO(2)_1$ factor corresponds to the fermions which are 
neutral in the process so they won't be considered 
afterwards.

\subsubsection{The gauge sector}
To construct the model we assume that 
the gauge sector of the heterotic strings contain an unbroken 
$SO(6)_1$, whose 
contribution to the partition function is,
written in terms of $SO(6)_1$ free fermionic characters
$\bar{\Xi}^{s_6}_6$, see App.~\ref{appLie}.
Since we decompose the characters of the left-moving 
sector according to eq.~(\ref{decompA3}), 
a natural choice for the action of the deformation in the 
right-moving gauge sector is to use a 
similar Kazama-Suzuki decomposition, but for 
$k=3$, in which case the bosonic \textsc{cft} is trivial:
\begin{equation}
SO(8)_1 \to 
\frac{SO(4)_1}{SU(2)_{1} \times 
U\left( 1 \right)_{9}} \times \frac{SU(2)_{1} \times SO(2)_1}{U\left( 1 \right)_{3}}
\times U\left( 1 \right)_{3} \times U\left( 1 \right)_1 \times SO(2)_1
\end{equation}
Since as quoted previously two fermions --~the 
$SO(2)_1$ factor~-- are neutral it is enough 
that the gauge sector contains an $SO(6)_1$ subgroup.
To achieve this decomposition, first we decompose the 
$SO(6)_1$ characters in terms of $SO(4)_1 \times SO(2)_1$:
\begin{equation}
\bar{\Xi}^{\bar{s}_6}_6 = \sum_{\bar{s}_4,\bar{s}_{2} \in \zi_4} 
C \left[ \bar{s}_6;\bar{s}_4,\bar{s}_{2} \right] 
\bar{\Xi}^{\bar{s}_4}_4 \bar{\Xi}^{\bar{s}_{2}}_2 
\end{equation}
where the coefficients of the 
decomposition $SO(6) \to SO(4) \times SO(2)$ are either zero or one. 
And then we perform a coset decomposition for the 
$SO(4)_1$ characters:
\begin{equation}
\bar{\Xi}^{\bar{s}_4}_4 = \sum_{\ell = 0,1} \sum_{u \in \zi_{18}} 
\bar{\varpi}^{\bar{s}_4}_{\ell \, u} \bar{\chi}^{\ell}  \ \frac{\bar{\Theta}_{u,9}}{\bar \eta}
\end{equation}
in terms of $SU(2)_1$ characters $\bar{\chi}^{\ell}$ and 
$U\left( 1 \right)$ characters $\bar{\Theta}_{u,9}$. 
It defines implicitely the coset characters 
$\bar{\varpi}^{\bar{s}_4}_{\ell \, u}$.
Then the $SU(2)_1 \times SO(2)_1$ characters are decomposed as:
\begin{equation}
\bar{\chi}^{\ell} \bar{\Xi}^{\bar{s}_{2}}_2 = \sum_{v \in \zi_6} 
\bar{\varpi}^{\ell,\bar{s}_{2}}_{v} \frac{\bar{\Theta}_{v,3}}{\bar \eta}\, ,
\end{equation}
So putting together these branching relations we have the 
following Kazama-Suzuki decomposition for the free fermions 
of the gauge sector:
\begin{equation}
\bar{\Xi}^{\bar{s}_6}_6 = \sum_{\bar{s}_4,\bar{s}_{2} \in 
\zi_4 } \sum_{\ell = 0,1} \sum_{u \in \zi_{18}} 
\sum_{v \in \zi_6} 
C \left[ \bar{s}_6;\bar{s}_4,\bar{s}_{2}\right] \ 
\bar{\varpi}^{\bar{s}_4}_{\ell \, u} \ 
\bar{\varpi}^{\ell,\bar{s}_{2}}_{v} \ 
\frac{\bar{\Theta}_{u,9}}{\bar \eta}\ \frac{\bar{\Theta}_{v,3}}{\bar \eta}.
\end{equation}

\subsubsection{The deformation}
Now we are in position to perform the asymmetric deformation 
adding a magnetic field to the model. 
The deformation acts on the following combination of 
left and right theta functions:
\begin{equation}
  \Theta_{n,3k} \, \bar{\Theta}_{u,9} \times 
  \Theta_{m,k} \bar{\Theta}_{v,3}.    
\end{equation}
As for the case of $SU(2)$~\cite{Israel:2004vv}, 
we have to assume that the level obeys the condition: 
\begin{equation}
\sqrt{\frac{k}{3}} = p \in \mathbb{N}\, , \end{equation}
to be able to reach the geometric coset point in the moduli 
space of \textsc{cft}. 
Then we have to perform $O(2,2,\mathbb{R})$ boosts in the lattices 
of the $U\left( 1 \right)$'s, mixing the left Cartan lattice 
of the super-\textsc{wzw} model with the right lattice 
of the gauge sector. These boosts are  
parameterized in function of the magnetic fields as:
\begin{equation}
\cosh \Omega_a = \frac{1}{1-2\textsc{h}_{a}^2}\ , \ \  
a = 1,2
\end{equation}
Explicitly we have:
\begin{multline}
  \sum_{N_1,N_2 \in \zi} q^{3k\left(N_1 + \frac{m}{6k}\right)^2}
  q^{k\left(N_2 + \frac{n}{2k}\right)^2} 
  \ \times 
  \sum_{f_1,f_2 \in \zi} \bar{q}^{9\left(f_1 + \frac{u}{18}\right)^2}
  \bar{q}^{3\left(f_2 + \frac{v}{6}\right)^2}\\
  \to 
  \sum_{N_1,N_2,f_1,f_2 \in \zi}
  q^{9 \left[ p \left( N_1 + \frac{m}{18p^2}\right) \cosh \Omega_1 
      + \left( f_1 + \frac{u}{18}\right) \sinh \Omega_1 \right]^2}
  q^{3 \left[ p \left( N_2 + \frac{n}{6p^2}\right) \cosh \Omega_2 
      + \left( f_2 + \frac{v}{6}\right) \sinh \Omega_2 \right]^2}\\
  \times \ 
  \bar{q}^{9 \left[ 
      \left( f_1 + \frac{u}{18}\right) \cosh \Omega_1
      + p \left( N_1 + \frac{m}{18p^2}\right) \sinh \Omega_1\right]^2}
  \bar{q}^{3 \left[ 
      \left( f_2 + \frac{v}{6}\right) \cosh \Omega_2 
      + p \left( N_2 + \frac{n}{6p^2}\right) \sinh \Omega_2 \right]^2}
\end{multline}
After an infinite deformation, we get the following constraints on the charges:
\begin{subequations}
  \begin{align}
    m &= p \left(18\mu - u \right), \ \mu \in \zi_p \\
    n &= p \left( 6\nu  - v \right), \ \nu \in \zi_p 
  \end{align}
\end{subequations}
and the $U\left( 1 \right)^2$ \textsc{cft} that has been deformed marginally decouples from the rest and 
can be safely removed.
In conclusion, the infinite deformation gives:
\begin{multline}
Z^{(s_4,s_2;\bar{s}_6 )}_{F_3} \left(\tau \right) = \sum_{\Lambda} \sum_{j}
  \sum_{\mu,\nu \in \zi_{p}}\ \sum_{\bar{s}_4,\bar{s}_{2} \in \zi_4 } C \left[
    \bar{s}_6;\bar{s}_4,\bar{s}_{2}\right]\\ \sum_{\ell = 0,1} \sum_{u \in \zi_{18}}\ 
  \sum_{v \in \zi_6}  
  \mathcal{C}^{\Lambda \, (s_4)}_{j\ , \ p (18\mu-u)} \ \mathcal{C}^{j\,
    (s_2)}_{p(6\nu -v) } \times \bar{\chi}^{\Lambda}_{k-3} \ \bar{\varpi}^{\bar{s}_4}_{4;\,
    \ell u} \ \bar{\varpi}^{\ell,\bar{s}_{2}}_{v}  
\end{multline} 
where the sum over $\Lambda, j$ runs over integrable representations, see 
appendix~~\ref{appLie}.
This is the partition function for the $SU\left(3\right)/U\left( 1 \right)^2$ coset space.  The
fermionic charges in the left and right sectors are summed according to the
standard rules of Gepner heterotic constructions~\cite{Gepner:1988qi}. The
modular properties of this partition function are the same as before the
deformation, concerning the $\zi_4$ indices of the worldsheet fermions.


\subsubsection{Alternative approach: direct abelian coset}

Here we would like to take a different approach, by deforming directly the
Cartan lattice of $\mathfrak{su}_3$ without decomposing the left \textsc{cft} 
in terms of \textsc{ks} $N=2$ theories.  As explained
in the App.~\ref{appLie}, it is possible to perform a generalized
(super)parafermionic decomposition of the characters of the
$\hat{\mathfrak{su}}_3$ super-algebra at level $k$ (containing 
a bosonic algebra at level $k-3$) w.r.t. the Cartan torus:
\begin{equation}
  \chi^{\Lambda}  \ 
  \left(
    \frac{\vartheta \oao{a}{b}}{\eta}\right)^{\mathrm{dim} (\mathfrak j ) /2}  
= 
  \sum_{\lambda \in \mathbf{M}^\ast \mathrm{mod} \ k \mathbf{M}} 
  \mathcal{P}^{\Lambda}_{\lambda} \oao{a}{b}  
  \frac{\Theta_{\lambda,k}}{\eta^{\mathrm{dim} (\mathfrak k )}}
\end{equation}
where the theta function of the $\widehat{\mathfrak{su}}_3$ 
affine algebra reads, 
for a generic weight $\lambda = m_i \lambda^{i}_f$ (see app.~\ref{appLie}):
\begin{equation}
  \Theta_{\lambda,k} = \sum_{\gamma \in \mathbf{M} 
    + \frac{\lambda}{k} }
  q^{\frac{k}{2} \kappa (\gamma,\gamma)} = 
  \sum_{N^1, N^2 \in \zi} 
  q^{\frac{k}{2} \| N^1 \alpha_1 + N^2 \alpha_2 
    + \frac{m_1 \lambda_{f}^1 + m_2 \lambda_{f}^2}{k} \|^2}
\end{equation}
To obtain an anomaly-free model it is natural to associate this model 
with an abelian coset decomposition of an $SU(3)_1$ current algebra 
made with free fermions of the gauge sector. Thus if the 
gauge group contains an $SU(3)_1$ unbroken factor their characters 
can be decomposed as:
\begin{equation}
\bar{\chi}^{\bar{\Lambda}} = \sum_{\bar{\lambda}= \bar{n}_i \lambda^{i}_f  
\ \in \ \mathbf{M}^\ast 
\mathrm{mod} \ \mathbf{M}} \bar{\varpi}^{\bar{\Lambda}}_{\bar{\lambda}}
\bar{\Theta}_{\bar{\lambda}}.
\end{equation}
Again we will perform the asymmetric deformation as a boost between 
the Cartan lattices of the left $\hat{\mathfrak{su}}_3$ algebra 
at level $k$ and the right $\hat{\mathfrak{su}}_3$ lattice algebra at level one 
coming from the gauge sector.
So after the infinite deformation we will get the quantization condition $\sqrt{k}=p$
and the constraint:
\begin{align}
\lambda + p \bar{\lambda} = 0 \mod p \, \textbf{M} = : p \, \mu \ , \ \ 
\mu \in \textbf{M}.
\end{align}
So we get a different result compared to the Kazama-Suzuki construction. It is
so because the constraints that we get at the critical point 
force the weight lattice of the $\hat{\mathfrak{su}}_3$ at level 
$k$ to be projected onto $p$ times the $\hat{\mathfrak{su}}_3$ weight lattice at level 
one of the fermions. This model does not correspond to a
K\"ahlerian manifold and should correspond to the $SU\left(3\right)$-invariant metric on
the flag space. Indeed with the \textsc{ks} method we get instead a projection onto 
$p$ times a lattice of $\hat{\mathfrak{su}}_3$ at level one 
which is dual to the orthogonal sublattice defined by $\alpha_1 \zi 
+ (\alpha_1 + 2 \alpha_2)\zi$--~in other words the lattice 
obtained with the Gell-Mann Cartan generators. 
In this case it is possible to decompose the model in \textsc{ks} cosets models 
with $N=2$ superconformal symmetry.\footnote{For the symmetrically 
gauged \textsc{wzw} models, this has been studied in~\cite{Eguchi:2003yy}.} 

We have seen in section~\ref{sec:gauging} that, in the gauging approach, ones
obtains the K\"ahler metric automatically when one starts from the $SU(2)$
fibration over $SU\left(3\right)/SU(2)$ rather than from the \textsc{wzw} model
$SU\left(3\right)$. It is now very easy to understand why it is the case. Indeed once
the $SU(2)$ has been taken out of the $SU\left(3\right)$, the only $U\left( 1 \right)$ 
that can be gauged (or deformed) is the $U\left( 1 \right)$ 
orthogonal to the root $\alpha_1$ of the
$SU(2)$ subalgebra, thus must be along the $\alpha_1 + 2\alpha_2$ vector. This will
allow automatically to decompose the abelian coset into \textsc{ks} Hermitean symmetric 
spaces, and the model corresponds to the K\"ahlerian metric on the flag space. 
However, at the level of the effective action, the deformation method of 
section~\ref{sec:su3-u1-u1} is not sensitive to these two possible \textsc{cft} 
realizations of the flag space.

\subsection{Generalization}
The previous construction can be easily generalized to any affine Lie algebra, 
but the formalism gets a little bit bulky. We will consider separately all the 
families of simple Lie algebras, since the construction differ significantly. 
We will mainly focus below on the \textsc{ks} decomposition method.

\subsubsection{\boldmath $A_n$ \unboldmath algebras}

For an $SU(n+1)$ \textsc{wzw} model we use the following decomposition 
in terms of $N=2$ Kazama-Suzuki models:
\begin{multline}
  SU(n+1)_{k-n-1} \times SO(n^2+2n)_1 \to \\ \frac{SU(n+1)_{k-n-1} \times SO(2n)_1
  }{SU(n)_{k-n} \times U\left( 1 \right)_{ n(n+1)k/2}} \times \frac{SU(n)_{k-n} \times
    SO(2(n-1))_1}{SU(n-1)_{k-n+1} \times U\left( 1 \right)_{(n-1)n k/2}} \times \cdots \times
  \frac{SU(2)_{k-2} \times SO(2)_1 }{U\left( 1 \right)_{k}} \\
   \times \ SO(n)_1 \times U\left( 1 \right)_{n(n+1)k/2} \times U\left( 1 \right)_{(n-1)n k/2} \times \cdots \times U\left( 1 \right)_k \\
\end{multline}
So the left worldsheet fermions of $SO(n^2+2n)_1$ are decomposed into:
\begin{equation}
  SO(n^2+2n) \to SO(2n)_1 \times 
  SO(2(n-1))_1 \times \cdots 
  \times SO(2)_1 \times SO(n)_1
\end{equation}
where $n$ fermions are neutral.  The Kazama-Suzuki decomposition of the
characters reads:
\begin{multline}
  \chi^{\Lambda} \ \Xi^{s_{2n}}_{2n} \Xi^{s_{2(n-1)}}_{2(n-1)} \cdots \Xi^{s_2}_2
  \Xi^{s_1}_n = \sum_{\Lambda^1,\Lambda^2,\ldots,j}\ 
  \sum_{m_1 \in \zi_{n(n+1)\, k}} \sum_{m_2 
    \in \zi_{{(n-1)n\, k}}} \cdots 
  \sum_{m_n \in \zi_k} \\
  \mathcal{C}^{\Lambda,\, (s_{2n})}_{\Lambda^1,\,m_1}
  \mathcal{C}^{\Lambda^1,\, (s_{2(n-1)})}_{\Lambda^2,\,m_2}
  \cdots \mathcal{C}^{j,\, (s_2)}_{m_n} 
  \ \Xi^{s_1}_n \ \times \ \frac{\Theta_{m_1, \frac{n(n+1)k}{2}}}{\eta}
  \frac{\Theta_{m_2, \frac{(n-1)n\, k}{2}}}{\eta} \ \cdots \ 
  \frac{\Theta_{m_n,k}}{\eta} 
  \label{decompKSAN}
\end{multline}
where the sum on $\Lambda^1,\Lambda^2,\ldots,j$ is taken over integrable representations
(see App.~\ref{appLie}) of $SU(n),SU(n-1),\ldots,SU(2)$.  For the right
fermions of the gauge sector the story is the same as for the $SU\left(3\right)$
example. We will need $n(n+1)$ free fermions realizing an $SO(n^2+n)_1$
algebra, in order to use the Kazama-Suzuki decomposition for the $A_n$ model
at level $k=n+1$, such that the bosonic part trivializes:
\begin{multline}
  SO(n^2+n)_1 \to \\
  \frac{SO(2n)_1 }{SU(n)_{1} \times U\left( 1 \right)_{ 
      \frac{n(n+1)^2}{2}}}
  \times \frac{SU(n)_{1} \times SO(2(n-1))_1}{SU(n-1)_{2} 
    \times U\left( 1 \right)_{\frac{(n-1)n(n+1)}{2}}} \times \cdots \times 
  \frac{SU(2)_{n-1} \times SO(2)_1 }{U\left( 1 \right)_{n+1}} \times \\ 
  \times U\left( 1 \right)_{\frac{n(n+1)^2}{2}} \times 
  U\left( 1 \right)_{\frac{(n-1)n (n+1)}{2}} \times \cdots \times U\left( 1 \right)_{n+1}\\ 
  \label{decompgaugAN}
\end{multline}
So we can write the decomposition in terms of coset characters as:
\begin{multline}
  \bar{\Xi}^{\bar{s}_{n(n+1)}}_{n(n+1)} = \!\!\!\!\!\!\!
  \sum_{\bar{s}_{2n}, \bar{s}_{2(n-1)} 
    \cdots \bar{s}_{2} \in \zi_4 } 
  C \left[ \bar{s}_{n(n+1)}; \bar{s}_{2n}, \bar{s}_{2(n-1)}, 
    \ldots, \bar{s}_{2} \right] 
  \sum_{u_1 \in \zi_{n(n+1)^2}}
  \sum_{u_2 \in \zi_{(n-1)n(n+1)}}
  \!\!\!\!\!\!\!\! \cdots \sum_{u_n \in \zi_{2(n+1)}} \\
  \bar{\varpi}^{\bar{s}}_{\bar{\Lambda}_{1},\, u_1}
  \bar{\varpi}^{\bar{\Lambda}_{1},\, \bar{s}_{2n} }_{\bar{\Lambda}_{2},u_2} \cdots 
  \bar{\varpi}^{j,\, \bar{s}_{2}}_{u_n} \ \times \ 
  \frac{\Theta_{u_1,\frac{n(n+1)^2}{2}}}{\bar \eta}\ 
  \frac{\Theta_{u_2,\frac{(n-1)n(n+1)}{2}}}{\bar \eta} \ \cdots \ 
  \frac{\Theta_{u_n,n+1}}{\bar \eta}
\end{multline}
For the left coset to exist one has to assume the following constraint on the 
level of the $A_n$ affine algebra:
\begin{equation}
\sqrt{\frac{k}{n+1}} = p \in \zi
\end{equation}
Then the decomposition can be carried out straightforwardly, by mixing the 
lattices of the holomorphic theta function for the decomposition~(\ref{decompKSAN}) 
and the decomposition~(\ref{decompgaugAN}). We get the following constraints:
\begin{equation}
  \begin{cases}
    m_1 =  p \left[ n(n+1)^2 \mu_1 - u_1 \right], & \mu_1 \in \zi_p\\
    m_2 =  p \left[ (n-1)n(n+1) \mu_2 - u_2 \right], & \mu_2 \in \zi_p \\
     \cdots \\
    m_n =  p \left[ 2(n+1) \mu_n - u_n \right], & \mu_n \in \zi_p
  \end{cases}
\end{equation}

So at the end we can remove the $U\left( 1 \right)^{n}$ free \textsc{cft} contribution and 
we get the following ``partition function'' for 
the $SU(n+1)/U\left( 1 \right)^{n}$ left coset, with $N=2$ worldsheet superconformal 
symmetry:
\begin{multline}
  Z^{(s_{2n},\ldots,s_2;\bar{s}_{n(n+1)} )}_{F_{n+1}} \left(\tau \right) =
  \sum_{\Lambda} \sum_{\Lambda^1,\Lambda^2,\ldots,j} \ \sum_{m_1 \in \zi_{n(n+1)\, k}} \sum_{m_2 \in
    \zi_{{(n-1)n\, k}}} \cdots
  \sum_{m_n \in \zi_k} \\
  \sum_{\bar{s}_{2n}, \bar{s}_{2(n-1)} \cdots \bar{s}_{2} \in \zi_4 } C \left[
    \bar{s}_{n(n+1)}; \bar{s}_{2n}, \bar{s}_{2(n-1)}, \ldots, \bar{s}_{2}
  \right] \sum_{u_1 \in \zi_{n(n+1)^2}} \sum_{u_2 \in \zi_{(n-1)n(n+1)}} \cdots
  \sum_{u_n \in \zi_{2(n+1)}} \\
  \sum_{\mu_1,\ldots,\mu_n \in \zi_p} \mathcal{C}^{\Lambda,\, (s_{2n})}_{\Lambda^1,\, p \left[
      n(n+1)^2 \mu_1 - u_1 \right]} \mathcal{C}^{\Lambda^1,\,
    (s_{2(n-1)})}_{\Lambda^2,\, p \left[ (n-1)n(n+1) \mu_2 - u_2 \right]} \cdots
  \mathcal{C}^{j,\, (s_2)}_{p \left[ 2(n+1) \mu_n - u_n \right]} \\ \ \times \  
  \bar{\chi}^{\Lambda} \bar{\varpi}^{\bar{s}}_{\bar{\Lambda}_{1},\, u_1}
  \bar{\varpi}^{\bar{\Lambda}_{1},\, \bar{s}_{2n} }_{\bar{\Lambda}_{2},u_2} \cdots
  \bar{\varpi}^{j,\, \bar{s}_{2}}_{u_n}
\end{multline}
As in the previous example this characters combination behaves
covariantly under modular transformation, \emph{i.e.} is modular invariant up to 
the transformation of the fermionic indices $\{ s_i \}$ and $\bar{s}_{n(n+1)}$. 
The modular invariance of the complete heterotic string background 
will be ensured by an appropriate Gepner construction. 

Now let us consider the other simple Lie algebras. For sake of brevity we will 
only sketch the method, which is quite parallel to the present case.

\boldmath
\subsubsection{B$_n$ algebras}
\unboldmath

In this case, the relevant Kazama-Suzuki $N=2$ coset model is:
\begin{equation}
  \frac{SO (2n+1)_{k-2n+1} \times SO(4n-2)_1}{
    SO(2n-1)_{k-2n+3} \times U\left( 1 \right)_{2k}}    
\end{equation}
therefore the decomposition in $N=2$ models of the group manifold is:
\begin{multline} 
  SO(2n+1)_{k-2n+1} \times SO(n(2n+1))_1 \to \\
  \to \frac{SO(2n+1)_{k-2n+1} \times SO(4n-2)_1 }{ SO(2n-1)_{k-2n+3} \times U\left( 1 \right)_{2k}} \times
  \frac{SO(2n-1)_{k-2n+3} \times SO(4n-6)_1 }{
    SO(2n-3)_{k-2n+5} \times U\left( 1 \right)_{2k}} \times \\
  \times \cdots \times \frac{SO\left(3\right)_{k-1} \times SO(2)_1 }{U\left( 1 \right)_{2k}} \times SO(n)_1 \times
  \left(U\left( 1 \right)_{2k}\right)^n
\end{multline}
So there are no specific constraints on the right fermions of the gauge 
sector. We only need to pick up $n$ complex fermions with arbitrary boundary conditions, 
realizing an $\left[ SO(2)_1 \right]^n$ algebra~\footnote{Of course this algebra 
may be enhanced in the specific model at hand but this is not necessary. 
Note also that there is another construction when ones starts with 
and $SO(2n^2)_1$ algebra in the gauge sector and decompose it in 
terms of the $B_n$ Kazama-Suzuki model at level $2n-1$.
}
The level of the $SO(2n+1)$ has to be quantized as 
$\sqrt{k} \in \mathbb{N}$. Under this condition the deformation 
can be carried out straightforwardly.

\boldmath
\subsubsection{C$_n$ algebras}
\unboldmath

We consider here the \textsc{ks} cosets:
\begin{equation}
\frac{Sp (2n)_{k-n-1} \times SO(n(n+1))_1}{
SU(n)_{2k-n} \times U\left( 1 \right)_{nk}}    
\end{equation}
So apart from the first step the decomposition follows the 
pattern for $A_n$ algebras:
\begin{footnotesize}
  \begin{multline}
    Sp (2n)_{k-n-1} \times SO(n(2n+1))_1 \to \\
    {\footnotesize  \to \frac{Sp (2n)_{k-n-1} \times SO(n(n+1))_1}{%
        SU(n)_{2k-n} \times U\left( 1 \right)_{nk}}%
      \times \frac{SU(n)_{2k-n} \times SO(2(n-1))_1}{SU(n-1)_{2k-n+1}%
        \times U\left( 1 \right)_{(n-1)n\, k}}%
      \times \cdots \times \frac{SU(2)_{2k-2} \times SO(2)_1 }{U\left( 1 \right)_{2k}} \times}
    \\ 
    \times SO(n)_1 \times U\left( 1 \right)_{nk} \times 
    U\left( 1 \right)_{(n-1)n\, k} \times U\left( 1 \right)_{(n-2)(n-1)\, k} \times \cdots 
    \times U\left( 1 \right)_{2k}
  \end{multline}
\end{footnotesize}
Then one need in the gauge sector an $SO(2n^2)_1$ algebra that will be split 
according to the purely fermionic Kazama-Suzuki decomposition 
for $C_n$, together with the quantization condition 
\begin{equation} \sqrt{\frac{k}{n+1}} \in \mathbb{N} \end{equation}
Then the deformation will lead to the flag space 
partition function.

\boldmath
\subsubsection{D$_n$ algebras}
\unboldmath

We consider here the \textsc{ks} cosets:
\begin{equation}
  \frac{SO (2n)_{k-2n+2} \times SO(n(n-1))_1}{
    SU(n)_{k-n} \times U\left( 1 \right)_{2nk}}    
\end{equation}
This case is very close to the last one. We have the decomposition:
\begin{footnotesize}
  \begin{multline}
    SO(2n)_{k-2n+2} \times SO(n(2n-1))_1 \to \\
    \to \frac{SO (2n)_{k-2n+2} \times SO(n(n-1))_1}{
      SU(n)_{k-n} \times U\left( 1 \right)_{2nk}}
    \times 
    \frac{SU(n)_{k-n} \times SO(2(n-1))_1}{SU(n-1)_{k-n+1} 
      \times U\left( 1 \right)_{(n-1)n k /2}}
    \times \cdots \times \frac{SU(2)_{k-2} \times SO(2)_1 }{U\left( 1 \right)_{k}} \times \\
    \times SO(n)_1 \times U\left( 1 \right)_{2nk} \times 
    U\left( 1 \right)_{(n-1)n k/2} \times U\left( 1 \right)_{(n-2)(n-1) k/2} \times \cdots 
    \times U\left( 1 \right)_{k}
  \end{multline}
\end{footnotesize}
So the fermions of the gauge sector have to realize an 
$SO[2n(n-1)]_1$ algebra, 
together with the quantization condition 
\begin{equation}
\sqrt{\frac{k}{2n-2}} 
\in \mathbb{N}
\end{equation}

\subsubsection{Exceptional algebras}
The two exceptional algebras leading to $N=2$ theories --~\emph{i.e.} 
giving Hermitian symmetric coset spaces~-- are $E_6$ and 
$E_7$.
In the first case, we have the decomposition:
\begin{footnotesize}
  \begin{multline}
    (E_6)_{k-12} \times SO(78)_1 \to \\
    \to \frac{(E_6)_{k-12} \times SO(32)_1}{SO(10)_k \times U\left( 1 \right)_{6k}}
    \times \frac{SO(10)_k \times SO(20)_1}{SU(5)_{k+3}
      \times U\left( 1 \right)_{10(k+8)}} 
    \times \frac{SU(5)_{k+3} \times SO(8)_1}{SU(4)_{k+4}\times 
      U\left( 1 \right)_{10(k+8)}} \times \cdots \times 
    \frac{SU(2)_{k+6} \times SO(2)_1}{U\left( 1 \right)_{k+8}} \times \\
    \times SO(6)_1 \times U\left( 1 \right)_{6k} \times 
    U\left( 1 \right)_{10(k+3)} \times U\left( 1 \right)_{10(k+8)}\times \cdots \times U\left( 1 \right)_{k+8}  
  \end{multline}
\end{footnotesize}
In this case we need fermions in the gauge sector realizing 
an $E_6$ algebra at level one, and will lead to the quantization 
condition $\sqrt{k/12} \in \mathbb{N}$. 
In the second case, we have the decomposition:
\begin{equation}
(E_7)_{k-18} \times SO(133)_1 \to 
\frac{(E_7)_{k-18} \times SO(54)_1}{(E_6)_k \times U\left( 1 \right)_{3k}}
\times  \frac{(E_6)_{k} \times SO(32)_1}{SO(10)_{k+12} \times U\left( 1 \right)_{6(k+12)}}
\times \cdots 
\end{equation}
and we see clearly that the conditions on the level we would get from the 
first $U\left( 1 \right)$ at level $3k$ and the other ones are generically incompatible. 
Thus one cannot construct a flag space \textsc{cft} for $E_7$ but only 
a coset by the maximal torus of the $E_6$ embedded in $E_7$.


\subsection{Kazama-Suzuki decomposition vs. abelian quotient}
In this section we would like to stress the ambiguity in defining 
an abelian coset of \textsc{wzw} models. We will consider the $A_n$ case 
in the discussion, although it's pretty much the same for the other 
classical Lie algebras.

An abelian super-coset $\nicefrac{G \times SO(\#\mathfrak{g} -d)}{U\left( 1 \right)^d}$, 
(with $\hat{\mathfrak{g}}$ at level $k-g^{\ast}$) must be 
supplemented with the definition of the action of the abelian subgroup in 
$g$, corresponding to a choice of a particular sub-lattice 
of $\Gamma  \in \sqrt{k} \mathbf{M}$ 
(these issues have been discussed in~\cite{Eguchi:2003yy} for symmetric 
supercosets of type II superstrings). 
In our construction, the left-coset structure will require that, 
in order to achieve modular invariance, the lattice behaves covariantly 
as some combination of right-moving fermions of the gauge sector 
of the heterotic string. It will be possible only if the level 
of the $\hat{\mathfrak g}$ affine algebra obeys a special 
quantization condition. In the \textsc{ks} construction we define 
with these right-moving fermions an orthogonal lattice; therefore we have 
also to choose an orthogonal sub-lattice of the root lattice 
for the \textsc{wzw} model in order to make this construction 
possible. 

For the $A_n$ algebra, the relevant orthogonal basis is written as 
follows:\footnote{In the case of $A_2$, we find the Gell-Mann matrices 
of $SU\left(3\right)$ \eqref{eq:Gell-Mann-matrices}.}
\begin{equation}
  \begin{cases}
    \nu_1 = \sqrt{k}\alpha_1 & \left(\nu_1 , \nu_1 \right) = 2 k\\
    \nu_2 = \sqrt{k} \left(\alpha_1 + 2 \alpha_2 \right) & \left(\nu_2 , \nu_2 \right) = 6 k\\
    \nu_3 = \sqrt{k} \left(\alpha_1 + 2 \alpha_2+3\alpha_3 \right)
    & \left(\nu_3 , \nu_3 \right) = 12 k\\
    \cdots \\
    \nu_n = \sqrt{k} \left(\alpha_1 + 2 \alpha_2+\cdots + n\alpha_n \right) & \left(\nu_n , \nu_n \right)
    = n \left( n+1 \right) k
  \end{cases}
\end{equation}
and is of course a sub-lattice of the complete root lattice. 
More precisely it corresponds to:
\begin{equation}
  \sqrt{k}\, \Gamma \ = \ \sqrt{k} \bigoplus_{a=1}^n 
  a \,  \zi \, \alpha_a \ \subset \ \sqrt{k} \bigoplus_{a = 1}^{n}
  \zi \, \alpha_a    
\end{equation}
Then the associated theta-functions of $\hat{\mathfrak{su}}_{n+1}$can be
written as a product of usual $\hat{\mathfrak{su}}_2$ theta functions:
\begin{equation}
  \Theta^{(\Gamma)}_{\lambda,k} = \prod_{a=1}^{n} \Theta_{m_a,a(a+1)k/2 } \quad 
  \text{with} \quad \lambda = \sum m_a \nu_{a}^\ast .    
\end{equation}

This choice of orthogonal basis allows actually to decompose 
the abelian coset into a chain of Kazama-Suzuki models, with enhanced
$N=2$ supersymmetry on the worldsheet. Indeed we have to choose
the lattice of the $U\left( 1 \right)$ in $\nicefrac{SU(n+1)}{SU(n) \times
  U\left( 1 \right)}$ to be $\zi \nu_n$, such that it will be orthogonal to
the root lattice of $\mathfrak{su}_{n-1}$ given by $\sum_{a=1}^{n-1} \zi
\alpha_a$, thus allowing to gauge it.
 
The left coset corresponding to this choice of abelian subgroup is obtained
by a marginal deformation with the operator $\bigoplus_a (\nu_a , \textsc{h})$. Its
partition function is composed of the coset characters obtained through the
branching relation:
\begin{equation}
  \chi^{\Lambda} \prod_{r=1}^{n^2+n} \Theta_{s_r,2} = \sum_{\lambda = \lambda_a \nu^{\ast}_a \ \in \ \Gamma^\ast
    \ \text{mod} \ k \Gamma} C^{\Lambda \, (s_1, \cdots, s_{n^2+n} )}_{\lambda} \prod_{a=1}^{n}
  \Theta_{\lambda_a, a(a+1)k/2}.    
\end{equation}

On the other hand, the standard $N=1$ abelian coset construction
is defined with a full $\sqrt{k}\mathbf{M}$ lattice. 
The left coset is obtained by a
marginal deformation with the operator $\oplus_a (\alpha_a , \textsc{h})$. The relevant coset
characters are given by:
\begin{equation}
  \chi^{\Lambda} \prod_{r=1}^{n^2+n} \Theta_{s_r,2} = 
  \sum_{\lambda = \in M^\ast \ \text{mod} \  k M} 
  \tilde{C}^{\Lambda \, (s_1, \cdots, s_{n^2+n} )}_{\lambda} 
  \Theta_{\lambda,k}.  
\end{equation}

As in the $A_3$ case, we can show that the \emph{left cosets} corresponding to these two 
classes of models are different. They are in correspondence with the 
different possible metrics (K\"ahlerian and non-K\"ahlerian) 
on asymmetric cosets spaces discussed in 
appendix~\ref{sec:coset-space-geometry}.


\setcounter{footnote}{0}

\section{New linear dilaton backgrounds of Heterotic strings}
\label{sec:heter-strings-sing}

These left-coset superconformal field theories can be used to construct 
various supersymmetric 
exact string backgrounds. The first class are generalizations of 
Gepner models~\cite{Gepner:1988qi} and Kazama-Suzuki 
constructions~\cite{Kazama:1989qp} using the left cosets as building 
blocks for the internal \textsc{scft}. This as already 
been considered in~\cite{Berglund:1996dv} for the $S^2$ coset 
but can be extended using the new theories constructed above. In this 
case there is no geometric interpretation from the sigma model point 
of view since these theories have no semi-classical limit. Indeed 
the levels of the cosets are frozen because their central charge 
must add up to $c=9$ (in the case of four-dimensional compactification). 
However we expect that they correspond to special points in the 
moduli spaces of supersymmetric compactifications, generalizing the 
Gepner points of the CY manifolds.

Another type of models are the left cosets analogues of the NS5-branes 
solutions~\cite{Callan:1991dj,Kounnas:1990ud} and of their extensions 
to more generic supersymmetric vacua with a dilaton background. It was shown 
in~\cite{Giveon:1999zm} that a large class of 
these linear dilaton theories are dual to singular 
CY manifolds in the decoupling limit. An extensive review 
of the different possibilities in various dimensions 
has been given in~\cite{Eguchi:2003yy} with all the 
possible $G/H$ cosets. 
The left cosets that we constructed allows to extend all these 
solutions to heterotic strings, with a different geometrical 
interpretation since our cosets differ from ordinary 
gauged \textsc{wzw} model. However the superconformal structure 
of the left sector of our models is exactly the same as for 
the corresponding gauged \textsc{wzw} --~except that 
the values of the N=2 R-charges that appear in the 
spectrum are constrained~-- so we can carry over all 
the known constructions to the case of the geometric 
cosets.

In the generic case these constructions 
involve non-abelian cosets, and as we showed the asymmetric 
deformations and gaugings apply only to the abelian components. 
Thus in general we will get mixed models which are gauged 
\textsc{wzw} models w.r.t. the non-abelian part of $H$ and 
geometric cosets w.r.t. the abelian components of $H$. 
Below we will focus on purely abelian examples, i.e. corresponding 
to geometric cosets. The dual interpretation of these models, 
in terms of the decoupling limit of some singular compactification 
manifolds, is not known. Note however that by construction there 
are about $\sqrt{k}$ times less massless states in our models 
than in the standard left-right symmetric solutions. Therefore 
they may correspond to some compactifications with fluxes, for which the 
number of moduli is reduced. It would be very interesting to investigate 
this issue further.

\paragraph{Six-dimensional model.}

We consider here the critical superstring background:
\begin{equation}
  \setR^{5,1} \times \frac{SL(2,\mathbb{R})_{k+2}\times SO(2)_1}{U\left( 1 \right)_{k}} 
  \times \left[ \sfrac{U\left( 1 \right)_k}{SU(2)_{k-2} \times SO(2)_1} \right]
\end{equation}
the second factor being a left coset \textsc{cft} as discussed in this paper. This 
is the direct analogue of the five-brane solution, or more precisely of the double 
scaling limit of NS5-branes on a circle~\cite{Giveon:1999px,IKPT}, in the present 
case with magnetic flux. This theory has $N=2$ charges but, in order to 
achieve spacetime supersymmetry one must project onto odd-integral 
$N=2$ charges on the left-moving side, 
as in the type II construction~\cite{IKPT}. This can be done 
in the standard way by orbifoldizing the left $N=2$ charges of the two cosets.

\paragraph{Four-dimensional model.}

A simple variation of the six-dimensional theory is given by 
\begin{multline}
\mathbb{R}^{3,1}\ \times\ \frac{SL(2,\mathbb{R})_{k/2+2}\times SO(2)_1}{U\left( 1 \right)_{2k}} \ \times 
\left[ \sfrac{U\left( 1 \right)_k}{SU(2)_{k-2} \times SO(2)_1} \right] \\
\times 
\left[ \sfrac{U\left( 1 \right)_k}{SU(2)_{k-2} \times SO(2)_1} \right] 
\end{multline}
which is the magnetic analogue of the 
(double scaling limit of) intersecting five-branes solution.
Also here an orbi\-foldization of the left $N=2$ charges is needed 
to achieve space-time supersymmetry.

\paragraph{Three-dimensional models: the flagbrane${}^\copyright$.}
 
We can construct the following background of the $G_2$ holonomy type, as in the 
case of symmetric coset~\cite{Eguchi:2001xa}:
\begin{equation}
  \setR^{2,1} \times  \mathbb{R}_Q  \times \left[  \sfrac{U\left( 1 \right)_k \times
      U\left( 1 \right)_{3k}}{SU\left(3\right)_{k-3} \times SO(6)_1} \right]
\end{equation}
and the non-trivial part of the metric is
\begin{equation}
  ds^2 = -\di t^2 + \di x^2 +  \di y^2 + \frac{k}{4r^2} 
  \left[ \di r^2 + 4 r^2 \di s^2 (\nicefrac{SU\left(3\right)}{U\left( 1 \right)^2}) \right].
\label{asyflagbrane}
\end{equation} 
Without the factor of four it would be a direct analogue of the NS5-brane, 
being conformal to a cone over the flag space. 

Another possibility in three dimensions is 
to lift the $SL(2,\mathbb{R})/U(1)$ coset to the group manifold 
$SL(2,\mathbb{R})$. In this case, as for the standard gauged \textsc{wzw} 
construction~\cite{Argurio:2000tg} we will get the following anti-de Sitter 
background:
\begin{equation}
SL(2,\mathbb{R})_{k/4+2}    \times \left[  
\sfrac{U\left( 1 \right)_{3k}}{SU\left(3\right)_{k-3} \times SO(6)_1} \right]
\end{equation}
and the left moving sector of this worldsheet \textsc{cft} defines 
an $N=3$ superconformal algebra in spacetime. 

\paragraph{Two-dimensional model}

In this case we can construct the background:
\begin{equation}
  \setR^{1,1} \times \frac{SL(2,\mathbb{R})_{k/4+2}\times SO(2)_1}{U\left( 1 \right)_{4k}} \times
  \frac{\sfrac{U\left( 1 \right)_{3k}}{SU\left(3\right)_{k-3} \times SO(6)_1}}{U\left( 1 \right)_k}  
\label{falseflagbrane}
\end{equation}
which corresponds in the classification of~\cite{Eguchi:2003yy} to a
non-compact manifold of $SU(4)$ holonomy once the proper projection is done
on the left $N=2$ charges. This solution can be also be thought as conformal to a cone
over the Einstein space $SU\left(3\right)/U\left( 1 \right)$.  
Using the same methods are for the NS5-branes in~\cite{IKPT}, we can show 
that the full solution corresponding to the model~(\ref{falseflagbrane}) can be obtained 
directly as the null super-coset:
\begin{equation}
\frac{SL(2,\mathbb{R})_{k/4} \times \sfrac{U\left( 1 \right)}{SU\left(3\right)_{k}}}{U\left( 1 \right)_L \times U\left( 1 \right)_R}
\end{equation}
where the action is along the elliptic generator in the $SL \left( 2,\setR
\right) $, with a normalization $\text{Tr} [(t^3 )^2 ] = -4$, and along the
direction $\alpha_1 + 2\alpha_2$ in the coset space $\ssfrac{U\left( 1
  \right)}{SU\left(3\right)}$, with a canonical normalization. 
For $r \to \infty$ the solution asymptotes the cone 
but when $r \to 0$ the strong
coupling region is smoothly capped by the cigar.

\setcounter{footnote}{0}
\section*{Acknowledgements}
We would like to thank J.~Troost for interesting discussions. 
The work of D.~I. has been supported by a 
Golda Meir Fellowship, the ISF Israel Academy of Sciences and
Humanities and the GIF German-Israel Binational Science Found. 

\newpage

\appendix
\newcommand{\D}[2]{{D^{#1}_{\phantom{#1}#2}}}
\newcommand{\K}[2]{{K^{#1}_{\phantom{#1}#2}}}
\setcounter{footnote}{0}
%


\section{Coset space geometry}
\label{sec:coset-space-geometry}

Coset spaces have been extensively studied in the mathematical literature of
the last fifty years. In this appendix we limit ourselves to collect some
classical results mainly dealing with the geometric interpretation. In
particular we will follow the notations of
\cite{Mueller-Hoissen:1988cq}.

Let $G$ be a semisimple Lie group and $H \in G $ a subgroup. As in the rest
of the paper, upper-case indices $\set{\textsc{m,n,o}}$ refer to the whole
group (algebra) $G$, lower-case indices $\set{m,n,o}$ to the subgroup
(subalgebra) and Greek indices $\set{\mu, \nu, \omega} $ to the coset.

It is useful to explicitly write down the commutation relations, separating the
generators of $H$ and $G/H$:
\begin{subequations}
  \begin{align}
    \comm{T_m, T_n} &= \F{o}{mn} T_o & \comm{T_m, T_\nu} &= \F{\omega}{m \nu} T_\omega
    \\
    \comm{T_\mu , T_\nu} &= \F{o}{\mu \nu } T_o + \F{\omega }{\mu \nu } T_\omega   
  \end{align}
\end{subequations}
Of course there are no $\F{\omega }{mn}$ terms since $H$ is a group.  $G/H$ is
said to be symmetric if $\F{\omega }{\mu \nu } \equiv 0$, \emph{i.e.} if the
commutator of any couple of coset elements lives in the dividing subgroup.
In this case a classical theorem states that the coset only admits one
left-invariant Riemann metric that is obtained as the restriction of the
Cartan-Killing metric defined on $G$ (see \emph{eg}~\cite{Kobayashi:1969}).
This is not the case when $H$ is the maximal torus (except for the most
simple case $G = SU \left( 2 \right)$) and the coset manifold accepts
different structures.

Any metric (or, more generally, any degree-2 covariant tensor) on $G/H$ can
be put in the form
\begin{equation}
  g = g_{\mu \nu} \left( x \right)J^\mu \otimes J^\nu  
\end{equation}
One can show that the $G$ invariance of $g$ is equivalent to:
\begin{equation}
  \label{eq:G-invariant-Metric}
  \F{\kappa}{a \mu } g_{\kappa \nu } \left( x \right)+
  \F{\kappa}{a \nu } g_{\kappa \mu } \left( x \right)= 0  
\end{equation}
and the homogeneity imposes
\begin{equation}
  g_{ij} = \text{constant}  
\end{equation}
Both conditions are easily satisfied by $g_{\mu \nu } \propto \delta_{\mu \nu}$ (this is the metric
on $G/H$ that we obtained in Eq.~\eqref{eq:G/TxT-metric}).  The Levi-Civita
connection 1-forms $\omega^\mu_{\phantom{\mu}\nu}$ of $g$ are determined by
\begin{subequations}
  \begin{gather}
    \di g_{\mu \nu} - \omega^\kappa_{\phantom{\kappa}\mu} g_{\kappa \nu} -
    \omega^\kappa_{\phantom{\kappa}\nu} g_{\kappa \mu} = 0\\
    \di J^\mu + \omega^\mu_{\phantom{\mu}\nu} \land J^\nu = 0 
  \end{gather}
\end{subequations}
and are explicitly written in terms of the structure constants as:
\begin{equation}
  \omega^\mu_{\phantom{\mu}\nu} = \F{\mu}{ a \nu} J^a +
  \D{\mu}{\rho \nu} J^\rho  
\end{equation}
where $\D{\mu}{\rho \nu}$ can be separated into its symmetric and antisymmetric
parts as follows:
\begin{subequations}
  \begin{align}
    \D{\mu}{\rho \nu} &= \frac{1}{2} \F{\mu}{ \rho \nu} +\K{\mu}{\rho \nu} \\
    \K{\mu}{\rho \nu} &= \frac{1}{2 } \left( g^{\mu \sigma }
      \F{\omega}{\sigma \rho} g_{ \omega \nu} + g^{\mu \sigma }
      \F{\omega}{\sigma \nu} g_{ \omega \rho}\right)
  \end{align}
\end{subequations}
We can then derive the curvature 2-form $\Omega = \di \omega + \omega \land
\omega $:
\begin{equation}
  \Omega^{\mu}_{\phantom{\mu}\nu} = \left( \D{\rho}{\sigma \nu} 
    \D{\mu}{\kappa \rho} -  \D{\rho}{\kappa \nu} \D{\mu}{\sigma \rho}
    -  \F{a}{\kappa\sigma} \F{\nu}{a \nu} -  \F{\rho}{\kappa \sigma}
    \D{\mu}{\rho \nu}\right) \frac{J^\kappa \land J^\sigma}{2}
\end{equation}
the Riemann tensor
\begin{multline}
  R^{\mu}_{\phantom{\mu}\nu \kappa \sigma} = 
  - \F{a}{\kappa \sigma} \F{\mu}{a \nu }   
  - \frac{1}{2}\F{\rho}{\kappa \sigma} \F{\mu}{\rho \nu }  
  + \frac{1}{4} \F{\rho}{\nu \kappa} \F{\mu}{\sigma \rho } 
  + \frac{1}{4} \F{\rho}{\nu \sigma} \F{\mu}{\kappa \rho } 
  + \frac{1}{2} \F{\rho}{\nu \kappa} \K{\mu}{\sigma \rho} 
  + \frac{1}{2} \F{\rho}{\nu \sigma} \K{\mu}{\kappa \rho} +\\
  - \frac{1}{2} \F{\mu}{\rho \kappa} \K{\rho}{\sigma \nu } 
  - \frac{1}{2} \F{\mu}{\rho \sigma} \K{\rho}{\kappa \nu } 
  - \F{\rho}{\kappa \sigma} \K{\mu}{\rho \nu } 
  + \K{\rho}{\sigma \nu} \K{\mu}{\kappa \rho } 
  - \K{\rho}{\kappa \nu} \K{\mu}{\sigma \rho }  
\end{multline}
and the Ricci tensor:
\begin{multline}
  Ric_{\nu \sigma} = R^{\mu}_{\phantom{\mu}\nu \mu \sigma} =
  - \F{a}{\mu \nu} \F{\mu}{a\nu} 
  - \frac{1}{2} \F{\rho}{\mu \sigma}\F{\mu}{\rho\nu}
  + \frac{1}{4} \F{\rho}{\nu \mu }\F{\mu}{\sigma\rho}
  + \frac{1}{2} \F{\rho}{\nu \mu}\K{\mu}{\sigma\rho}
  + \frac{1}{2} \F{\mu}{\rho \sigma}\K{\rho}{\mu \nu}+\\
  - \frac{1}{2} \F{\rho}{\mu \sigma}\K{\mu}{\rho\nu}
  - \frac{1}{2} \K{\rho}{\mu \nu}\K{\mu}{\sigma\rho}
\end{multline}

In particular, in the case of $g_{\mu \nu} = \delta_{\mu \nu}$ the expressions are greatly
simplified because the antisymmetric part $\K{\mu}{\nu\rho}$ vanishes and
then the Riemann and Ricci tensors are respectively given by:
\begin{align}
  R^{\mu}_{\phantom{\mu}\nu \kappa \sigma} &= - \F{a}{\kappa \sigma}
  \F{\mu}{a \nu } - \frac{1}{2}\F{\rho}{\kappa \sigma} \F{\mu}{\rho \nu } +
  \frac{1}{4} \F{\rho}{\nu \kappa} \F{\mu}{\sigma \rho } + \frac{1}{4}
  \F{\rho}{\nu \sigma} \F{\mu}{\kappa \rho } \\
  Ric_{\nu \sigma} &= - \F{a}{\mu \sigma } \F{\mu}{a\nu} 
  - \frac{1}{4} \F{\rho}{\mu \sigma}\F{\mu}{\rho\nu}  \label{eq:coset-Ricci}
\end{align}

Another fact that we used in the paper about $G/H$ cosets is a construction
due to Borel \cite{Borel:1958ch,Perelomov:1987va} of a K\"ahler structure over
$G/T$ where $T$ is the maximal torus. First of all we remark that such a
coset can be given a $\setC$ structure when associating holomorphic and
anti-holomorphic sectors to positive and negative roots respectively. One
can then show that the $\left( 1, 1\right)$ form defined as:
\begin{equation}
  \label{eq:two-form}
  \omega = \frac{\imath }{2} \sum_{\alpha >0} c_\alpha \mJ^\alpha \land \mJ^{\bar \alpha }  
\end{equation}
is closed if and only if for each subset of roots $\set{\alpha, \beta, \gamma} $ such
as $\alpha = \beta + \gamma $, the
corresponding real coefficients $c_\alpha $ satisfy the condition $c_\alpha = c_\beta +
c_\gamma $. Of course this is equivalent to say that the tensor
\begin{equation}
  \label{eq:coset-Kaehler-metric}
  g = \sum_{\alpha >0} c_\alpha \mJ^{\alpha } \otimes \mJ^{\bar \alpha}  
\end{equation}
is a K\"ahler metric on $G/T$.

\bigskip 

  
In particular, if we consider the $SU \left( 3 \right)$ group, for the
$\mathfrak{su} \left( 3 \right)$ algebra we can choose the Gell-Mann
$\lambda $ matrices \eqref{eq:Gell-Mann-matrices} as a basis. In this case
if we divide by the $U \left( 1 \right) \times U \left( 1 \right)$ subgroup
generated by $\braket{\lambda_3, \lambda_8}$, the most general metric
satisfying \eqref{eq:G-invariant-Metric} has the form $g = \diag
\set{a,a,b,b,c,c}$ \emph{ie} $SU \left( 3 \right)/ U\left( 1\right)\times
U\left( 1\right)$ admits a three parameter family of metrics. Among them,
the moduli space lines $a = b = c$ (the metric obtained in
Sec.~\ref{sec:su3-u1-u1}) and $a = b = c/2$ (the metric in
Sec.~\ref{sec:su-3-}) represent Einstein structures (with Ricci scalar
$15/a$ and $12/a$ respectively).  In both cases the manifold can be endowed
with complex structures (positive and negative roots respectively generating
the holomorphic and anti-holomorphic sectors) but only the latter admits a
K\"ahler structure (in this way we obtain the so-called flag space $F_3$).

\section{Equations of motion}
\label{sec:appendix-motion}

\subsection{Explicit derivation of some terms}
\label{sec:expl-deriv-some}

In this appendix we explicitly derive the expressions for the ${F^a}_{\mu \rho}
{F^a}_\nu^{\phantom{\nu}\rho}$ and $ H_{\mu \rho \sigma} H_{\nu}^{\phantom{\nu} \rho\sigma}$
terms appearing in the equations of motion \eqref{eq:beta-g}.

\paragraph{Gauge field strength.}
Consider the term coming from the gauge field strength.  First of all we can
build an orthonormal basis out of the Weyl-Cartan basis by complexifying the
Cartan generators and combining opposite ladder operators as follows:
\begin{equation}
  \label{eq:orthonormal-basis}
  \begin{cases}
    T^a = \imath H^a \\
    T^{2\mu - 1} = \imath \frac{\abs{\alpha_\mu }}{2} \left(
      E^{\alpha_\mu } - E^{- \alpha_\mu }
    \right) \\
    T^{2\mu } = \frac{\abs{\alpha_\mu }}{2} \left( E^{\alpha_\mu } +
      E^{- \alpha_\mu } \right)
  \end{cases}
\end{equation}
if we write explicitly the $\left( F^2\right)_{\mu \nu }$ term as
follows:
\begin{equation}
  \label{eq:F2-term}
  \left( F^2\right)_{\mu \nu } \propto \sum_{m,\omega} f^m_{\phantom{m}\nu \omega } f^m_{\phantom{m}\pi \omega  } = \sum_{m,\omega } \kappa \left(
    T^m, \left[ T^\nu, T^\omega \right] \right) \kappa \left(
    T^m, \left[ T^\pi , T^\omega \right] \right)
\end{equation}
we can see why rewriting everything this choice of basis simplifies the
calculation: the only commutators that will give a non-vanishing result when
projected on the Cartan generators are the ones involving opposite ladder
operators\footnote{We remember that the Cartan-Weyl basis is defined by:
  \begin{subequations}
      \label{eq:CartanWeyl}
    \begin{align}
      \left[ H^m, H^n \right] &= 0 \\
      \left[ H^a, E^{\alpha_\mu }\right] &= \left. \alpha_\mu \right|^a E^{\alpha_\mu } \\
      \left[ E^{\alpha_\mu }, E^{\alpha_\nu  }\right] &= 
      \begin{cases}
        N_{\mu,\nu } E^{\alpha_\mu + \alpha_\nu } & \text{if } \alpha_\mu + \alpha_\nu \in \Delta \\
        \frac{2}{\abs{\alpha_\mu }^2} \alpha_\mu \cdot H & \text{if } \alpha_\mu = -\alpha_\nu \\
        0 & \text{otherwise}
      \end{cases}
    \end{align}
  \end{subequations}},
that is $\left[ T^{2\mu-1}, T^{2 \mu } \right]$ which are explicitly given
by:
\begin{equation}
  \left[ T^{2\mu -1}, T^{2 \mu } \right] = \imath \frac{\abs{\alpha_\mu}^2}{4} 2 \left[
    E^{\alpha_\mu }, E^{-\alpha_\mu } \right] =  \alpha_\mu \cdot \left( \imath H\right)
\end{equation}
this means that:
\begin{equation}
  \begin{cases}
    \kappa \left( T^m, \left[ T^\nu, T^\omega \right] \right) = \left. \alpha_{\mu }
    \right|^m \delta_{\nu+1,\omega} & \text{if } \nu = 2 \mu - 1 \\
    \kappa \left( T^m, \left[ T^\nu, T^\omega \right] \right) = - \left. \alpha_{\mu }
    \right|^m \delta_{\nu-1,\omega} & \text{if } \nu = 2 \mu 
  \end{cases}
\end{equation}
putting this back in Eq.~\eqref{eq:F2-term} we find:
\begin{equation}
  \label{eq:ff-general}
  \sum_{m,\omega} f^m_{\phantom{m}\nu \omega } f^m_{\phantom{m}\pi \omega  } =
  \delta_{\nu \pi} 
  \begin{cases}
    \abs{\alpha_{\nicefrac{(\nu + 1)}{2}}}^2
       & \text{if $\nu$ is odd}\\
    \abs{\alpha_{\nicefrac{\nu }{2}}}^2
       & \text{if $\nu$ is even}
  \end{cases}
\end{equation}
if $\mathfrak{g}$ is simply laced then we can fix the normalizations to
$\abs{\alpha_{\mu }}^2 = \psi^2 \equiv2 $ and the above expression is greatly
simplified:
\begin{equation}
  \label{eq:ff-simply}
  \sum_{m,\omega} f^m_{\phantom{m}\nu \omega } f^m_{\phantom{m}\pi \omega  } =
  2 \delta_{\nu \pi }
\end{equation}
and by applying the right normalizations (see Eq.~\eqref{eq:bkd-fields}) we find
that for a general algebra:
\begin{equation}
  \label{eq:F2-term-general}
  {F^m}_{\nu \omega } g^{\omega \varpi} {F^m}_{\pi \varpi  } =
  \frac{4}{k_g}  \delta_{\nu \pi} 
  \begin{cases}
    \abs{\alpha_{\nicefrac{(\nu + 1)}{2}}}^2
       & \text{if $\nu$ is odd}\\
    \abs{\alpha_{\nicefrac{\nu }{2}}}^2
       & \text{if $\nu$ is even}
  \end{cases}
\end{equation}
and for a simply laced one:
\begin{equation}
  \label{eq:F2-term-simply}
  {F^m}_{\nu \omega } g^{\omega \varpi} {F^m}_{\pi \varpi  } = \frac{8}{k_g} \delta_{\nu \pi}
\end{equation}

\paragraph{NS-NS flux.}

From the definition of Casimir of the algebra we easily derive that:
\begin{equation}
  Q = - \sum_{\textsc{m}} \sum_{\textsc{o}} f^{\textsc{m}}_{\phantom{\textsc{m}}\textsc{no}} f^{\textsc{m}}_{\phantom{\textsc{m}}\textsc{op}} 
  = 2 g^\ast \delta_{\textsc{np}}
\end{equation}
where $g^\ast$ is the dual Coxeter number. Limit \textsc{n} and \textsc{p} to $\mathfrak{j}$
(and call them $\nu $ and $\pi$) and separate the two sums (that span over
the entire algebra) into the components over $\mathfrak{j}$ and $\mathfrak{k}$:
\begin{equation}
  \sum_{m \in \mathfrak{k}} \left( \sum_{o \in \mathfrak{k}} f^m_{\phantom{m}\nu o} 
    f^m_{\phantom{m}o\pi } +
    \sum_{\omega \in \mathfrak{j}} f^m_{\phantom{m}\nu \omega } f^m_{\phantom{m}\omega \pi }
  \right)
  + \sum_{\mu \in \mathfrak{j}} \left( \sum_{o \in \mathfrak{k}} f^\mu_{\phantom{\mu}\nu o}
    f^\mu_{\phantom{\mu}o\pi } +
    \sum_{\omega \in \mathfrak{j}} f^\mu_{\phantom{\mu}\nu \omega } f^\mu_{\phantom{\mu}\omega
      \pi } \right)= - 2 g^\ast \delta_{\nu \pi }
\end{equation}
now,
\begin{itemize}
\item the term with two elements in the Cartan is identically vanishing
  $f^m_{\phantom{m}\nu o} \equiv 0$ (for two generators in $\mathfrak{k}$ always
  commute)
\item the terms with one component in $\mathfrak{k}$ can be collected an
  interpreted as field strengths:
  \begin{equation}
    \sum_{m, \omega } f^m_{\phantom{m}\nu \omega } f^m_{\phantom{m}\omega \pi } +
    \sum_{o,\mu } f^\mu_{\phantom{\mu}\nu o} f^\mu_{\phantom{\mu}o\pi } 
  \end{equation}
\end{itemize}
and at the end of the day
\begin{equation}
  \sum_{\mu, \omega} f_{\nu \mu  \omega } f_{\pi \mu \omega } = 2 g^\ast 
  \delta_{\nu \pi} - 2 \sum_{m,\omega} f^m_{\phantom{m}\nu \omega }
  f^m_{\phantom{m}\pi \omega  } 
\end{equation}
so that for a general algebra, using \eqref{eq:ff-general}:
\begin{equation}
  \sum_{\mu, \omega} f_{\nu \mu  \omega } f_{\pi \mu \omega } = 2 g^\ast \delta_{\nu \pi} - 2
  \delta_{\nu \pi }
  \begin{cases}
    \abs{\alpha_{\nicefrac{(\nu + 1)}{2}}}^2
    & \text{if $\nu$ is odd}\\
    \abs{\alpha_{\nicefrac{\nu }{2}}}
    & \text{if $\nu$ is even}
  \end{cases}
\end{equation}
that reduces in the simply laced case to:
\begin{equation}
  \sum_{\mu, \omega} f_{\nu \mu  \omega } f_{\pi \mu \omega } = 2 \left( g^\ast - 2 \right)
  \delta_{\nu \pi}
\end{equation}
and with the proper normalizations:
\begin{equation}
  \label{eq:H2-term-general}
  H_{\nu \mu  \omega }g^{\mu\nu } g^{\omega\varpi} H_{\pi \nu \varpi } = 
  2 g^\ast \delta_{\nu \pi} - 2  \delta_{\nu \pi }
  \begin{cases}
    \abs{\alpha_{\nicefrac{(\nu + 1)}{2}}}^2
    & \text{if $\nu$ is odd}\\
    \abs{\alpha_{\nicefrac{\nu }{2}}}^2
    & \text{if $\nu$ is even}
  \end{cases}
\end{equation}
which reads in the simply laced case:
\begin{equation}
  \label{eq:H2-term-simply}
  H_{\nu \mu  \omega }g^{\mu \nu } g^{\omega\varpi} H_{\pi \nu \varpi } =  2 \left( g^\ast - 2 \right)
  \delta_{\nu \pi}
\end{equation}


\subsection{Equations of motion for the $F_3$ flag space}
\label{sec:flag-equations-motion}

To verify that the background fields that we obtained in
Sec.~\ref{sec:su-3-} solve the equations of motion at first order in $\alpha^\prime$
it is convenient to consider the complex structure defined on the $SU \left(
  3\right) / U \left( 1 \right)^2 $ coset by considering positive and
negative roots as holomorphic and anti-holomorphic generators respectively.

To fix the notation let the two simple roots be:
\begin{align}
  \alpha_1 = \comm{ \sqrt{2}, 0 }  && \alpha_2 = \comm{ - \nicefrac{1}{\sqrt{2}},
    \sqrt{\nicefrac{3}{2}} }
\end{align}
and the third positive root $\alpha_3 = \alpha_1 + \alpha_2 = \comm{
  \nicefrac{1}{\sqrt{2}}, \sqrt{\nicefrac{3}{2}}}$. We already know
that in the complex formalism the metric is diagonal and the coefficient
relative to the non-simple root is given by the sum of the two others as in
Eq.~\eqref{eq:coset-Kaehler-metric}. With the right normalization we have
the following metric and Ricci tensor:
\begin{align}
  g_{\mu \bar \nu } = \frac{k}{2} 
  \begin{pmatrix}
    1 \\
    & 1 \\
    & & 2
  \end{pmatrix} && R_{\mu \bar \nu } = 
  \begin{pmatrix}
    2 \\
    & 2 \\
    & & 4
  \end{pmatrix}
\end{align}

To write the structure constants we just have to remember the defining
relations for the Cartan--Weyl basis Eq.~\eqref{eq:CartanWeyl}: it is
immediate to see that $\F{1}{\mu \nu }$ and $\F{2}{\mu \nu }$ are
non-vanishing only if $\alpha_\mu $ and $\alpha_\nu $ are opposite roots
(which means in turn that in our complex formalism they are represented by
diagonal matrices) and, given the above choice of roots, we have:
\begin{align}
  \F{1}{\mu \bar \nu } = 
  \begin{pmatrix}
    \sqrt{2} \\
    & -\nicefrac{1}{\sqrt{2}} \\
    & & \nicefrac{1}{\sqrt{2}}
  \end{pmatrix} &&   \F{2}{\mu \bar \nu } = 
  \begin{pmatrix}
    0 \\
    & \sqrt{\nicefrac{3}{2}} \\
    & & \sqrt{\nicefrac{3}{2}}
  \end{pmatrix}
\end{align}
Let us now introduce a new tensor $C$ that in this basis assumes the form of
the unit matrix (this is indeed shown to be a tensor in
App.~\ref{sec:coset-space-geometry}):
\begin{equation}
  C_{\mu \bar \nu } = 
  \begin{pmatrix}
    1 \\
    & 1 \\
    & & 1
  \end{pmatrix} 
\end{equation}
we can use this tensor to define the $U \left( 1 \right)^2 $ gauge field
that supports the $F_3 $ background as\footnote{One can read this additional
  term with respect to the expression in Eq.~\eqref{eq:def-F-field} as a way
  to keep track of the fact that the embedded $SU \left( 2 \right)$
  subalgebra is at a different level with respect to the remaining currents.
  Actually this expression can be seen just as a generalisation of the
  initial one where we were restricting to cosets in which the currents
  played the r\^ole of vielbeins, \emph{i.e.} in this formalism the metric was
  proportional to the unit matrix.}:
\begin{equation}
  \FF{a}{\mu \bar \nu } = \sqrt{\frac{k}{2 k_g }} \F{a}{\mu \bar \rho } C^{\bar \rho \sigma
  } R_{\sigma \bar \nu }
\end{equation}

The only non-trivial equation of motion is $\beta^G = 0$ \eqref{eq:beta-g}:
\begin{equation}
  \beta^G = R_{\mu \bar \nu } - \frac{k_g}{4} \FF{a}{\mu \bar \sigma } g^{\bar \sigma  \rho } \FF{a
  }{\rho \bar \nu }
\end{equation}
in our basis all the tensors are diagonal matrices. For this reason it is
useful to pass to matrix notation. Let
\begin{equation}
  \mathbf{G} = 
  \begin{pmatrix}
    1 \\ 
    & 1 \\
    & & 2
  \end{pmatrix}
\end{equation}
so that the metric and the Ricci tensor are given by $\mathbf{g} =
\frac{k}{2} \mathbf{G}$ and $\mathbf{R } = 2 \mathbf{G}$. In this notation
the above equation reads:
\begin{multline}
  \beta^G = \mathbf{R} - \frac{k_g}{4}  \sum_{a=1}^2 \sqrt{\frac{k}{2 k_g}} \mathbf{f}^a
  \mathbf{R} \mathbf{g}^{-1} \sqrt{\frac{k}{2 k_g}} \mathbf{f}^a
  \mathbf{R} = \mathbf{R} - \frac{k}{8} \sum_{a=1}^2 \mathbf{f}^a \left( 2
    \mathbf{G} \right)
  \left(\frac{2}{k} \mathbf{G}^{-1} \right)  \mathbf{f}^a  \mathbf{R} =\\= \mathbf{R} -
  \frac{1}{2} \sum_{a=1}^2 \mathbf{f}^a \mathbf{f}^a \mathbf{R} = 0
\end{multline}
since $\sum_{a=1}^2 \mathbf{f}^a \mathbf{f}^a = 2 \uni{3} $ as one can see by
direct inspection.

\section{The \boldmath $SU\left( 3 \right)$ \unboldmath group: an explicit
  parametrization}
\label{sec:suleft-3right}

In this section we summarize some known facts about the representation of
the $SU \left( 3 \right)$ group so to get a
consistent set of conventions.

To obtain the the Cartan-Weyl basis $\set{H_a, E^{\alpha_j}}$ (defined in
Eq.~\eqref{eq:CartanWeyl}) for the $\mathfrak{su} \left( 3\right)$ algebra
we need to choose the positive roots as follows:
\begin{align}
  \alpha_1 = \comm{ \sqrt{2}, 0 } && \alpha_2 = \comm{ - \nicefrac{1}{\sqrt{2}},
    \sqrt{\nicefrac{3}{2}} } && \alpha_3 = \comm{ \nicefrac{1}{\sqrt{2}},
    \sqrt{\nicefrac{3}{2}}}
\end{align}
\begin{figure}[htbp]
  \begin{center}
     \includegraphics[width=4.5cm]{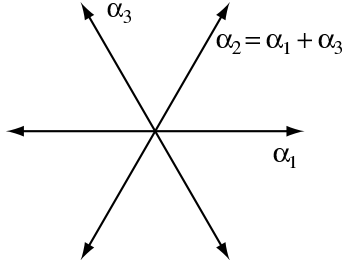}
  \end{center}
  \caption{Root system for $\mathfrak{su}\left( 3\right)$.}
  \label{fig:Root-sp4}
\end{figure}

The usual choice for the defining representation is:
\begin{small}
  \begin{align}
    H_1 = 
    \frac{1}{\sqrt{2}}\begin{pmatrix}
      1 & 0 & 0 \\
      0 & -1 & 0 \\
      0 & 0 & 0
    \end{pmatrix}&& H_2 = 
    \frac{1}{\sqrt{6}}\begin{pmatrix}
      1 & 0 & 0 \\
      0 & 1 & 0 \\
      0 & 0 & -2
    \end{pmatrix}&&
    E_1^+ = 
    \begin{pmatrix}
      0 & 1 & 0 \\
      0 & 0 & 0 \\
      0 & 0 & 0
    \end{pmatrix} && E_2^+ = 
    \begin{pmatrix}
      0 & 0 & 0 \\
      0 & 0 & 1 \\
      0 & 0 & 0
    \end{pmatrix}&& E_3^+ = 
    \begin{pmatrix}
      0 & 0 & 1 \\
      0 & 0 & 0 \\
      0 & 0 & 0
    \end{pmatrix} 
  \end{align}
\end{small}
and $E^-_j = \left( E_j^+ \right)^t$. 

A good parametrisation for the $SU \left( 3\right)$ group can be obtained
via the Gauss decomposition: every matrix $g \in SU \left( 3 \right)$ is
written as the product:
\begin{equation}
  g = b_- d b_+  
\end{equation}
where $b_-$ is a lower triangular matrix with unit diagonal elements, $b_+$
is a upper triangular matrix with unit diagonal elements and $d$ is a
diagonal matrix with unit determinant. The element $g$ is written as:
\begin{multline}
  g \left( z_1, z_2, z_3, \psi_1, \psi_2 \right) = \exp \left( z_1 E_1^- +
    z_2 E_3^- + \left( z_3 - \frac{z_1 z_2}{2}\right)E_2^- \right) \exp
  \left( -F_1 H_1 - F_2 H_2 \right)\\ \exp \left( \bar w_1 E_1^+ +
   \bar w_2 E_3^+ + \left( \bar w_3 - \frac{ \bar w_1 \bar
        w_2}{2}\right) E_2^+ \right) \exp \left( \imath \psi_1 H_1 + \imath \psi_2
    H_2 \right)
\end{multline}
where $z_\mu $ are 3 complex parameters, $\psi_i$ are two real and $F_1 $ and
$F_2 $ are positive real functions of the $z_\mu$'s:
\begin{equation}
  \begin{cases}
  F_1 = \log f_1 = \log \left( 1 + \abs{z_1}^2 + \abs{z_3}^2 \right)\\
  F_2 = \log f_2 = \log \left( 1 + \abs{z_2}^2 + \abs{z_3 - z_1 z_2 }^2 \right)
  \end{cases}
\end{equation}
By imposing $g \left( z_\mu , \psi_a  \right)$ to be unitary we find that the $w_\mu
$'s are complex functions of the $z_\mu$'s:
\begin{equation}
  \begin{cases}
    w_1 = - \frac{z_1 + \bar z_2 z_3}{\sqrt{f2}} \\
    w_2 = \frac{\bar z_1 z_3 - z_2 \left( 1+ \abs{z_1}^2
      \right)}{\sqrt{f_1}}\\
    w_3 = - \left( z_3 - z_1 z_2 \right)\sqrt{\frac{f_1}{f_2}}
  \end{cases}
\end{equation}
and the defining element $g \left( z_\mu, \psi_a \right)$ can then be written
explicitly as:
\begin{equation}
  \label{eq:5}
  g \left( z_1, z_2, z_3, \psi_1, \psi_2 \right) = 
  \begin{pmatrix}
    1 & 0 & 0 \\
    z_1 & 1 & 0 \\
    z_3 & z_2 & 1
  \end{pmatrix}
  \begin{pmatrix}
    \frac{1}{\sqrt{f_1}} & 0 & 0 \\
    0 & \sqrt{\nicefrac{f_1}{f_2}} & 0 \\
    0 & 0 & \sqrt{f_2}
  \end{pmatrix}
  \begin{pmatrix}
    1 & \bar w_1 & \bar w_3 \\
    0 & 1 & \bar w_2 \\
    0 & 0 & 1
  \end{pmatrix}
  \begin{pmatrix}
    e^{\imath \psi_1/2} & 0 & 0 \\
    0 & e^{-\imath \nicefrac{\left( \psi_1 - \psi_2 \right)}{2}}& 0\\
    0 & 0 & e^{\imath \nicefrac{\psi_2}{2}}
  \end{pmatrix}
\end{equation}
Now, to build a metric for the tangent space to $SU \left( 3 \right)$ we can
define the 1-form $\Omega\left( \mathbf{z}, \mathbf{\psi} \right) = g^{-1} \left(
  \mathbf{z }, \mathbf{\psi }\right) \di g \left( \mathbf{z }, \mathbf{\psi
  }\right)$ and write the Killing-Cartan metric tensor as $g_{\textsc{kc}} =
\tr \left( \Omega^\dag{} \Omega \right) = - \tr \left( \Omega \Omega \right)$ where we have used
explicitly the property of anti-Hermiticity of $\Omega $ (that lives in the
$\mathfrak{su} \left( 3 \right)$ algebra). The explicit calculation is lengthy
but straightforward.  The main advantage of this parametrization from our
point of view is that it allows for a ``natural'' embedding of the $SU
\left( 3 \right) / U\left( 1\right)^2 $ coset (see \emph{e.g.}
\cite{Gnutzmann:1998JP} or \cite{Kondo:1999tj}): in fact in these
coordinates the K\"ahler potential is
\begin{equation}
\label{eq:Kahler-SU3}
  K \left( z_\mu, \bar z_\mu  \right) = \log \left( f_1\left( z_\mu \right)
    f_2\left( z_\mu \right) \right) = \log \left[ \left( 1 +
      \abs{z_1}^2 + 
      \abs{z_3}^2 \right) \left( 1 + \abs{z_2}^2 + \abs{z_3 - z_1 z_2
      }^2 \right) \right]  
\end{equation}
and the coset K\"ahler metric is hence simply obtained as:
\begin{equation}
  g_{\alpha \bar \beta} \ \di z^\alpha \otimes \di \bar z^\beta  = \frac{\partial^2}{\partial z_\alpha \partial \bar z_\beta} K \left( z_\mu,
    \bar z_\mu  \right) \di z^\alpha \otimes \di \bar z^\beta   
\end{equation}

Another commonly used $\mathfrak{su} \left( 3 \right)$ basis is given by the
Gell-Mann matrices:
\begin{small}
  \begin{equation}
    \label{eq:Gell-Mann-matrices}
    \begin{array}{cccccccc}
      \gamma_1 &= 
      \frac{1}{\sqrt{2}}\begin{pmatrix}
        0 & \imath & 0 \\
        \imath & 0 & 0 \\
        0 & 0 & 0
      \end{pmatrix} &  \gamma_2 &= 
      \frac{1}{\sqrt{2}}\begin{pmatrix}
        0 & 1 & 0 \\
        -1 & 0 & 0 \\
        0 & 0 & 0
      \end{pmatrix} &  \gamma_3 &= 
      \frac{1}{\sqrt{2}}\begin{pmatrix}
        \imath  & 0 & 0 \\
        0 & -\imath  & 0 \\
        0 & 0 & 0
      \end{pmatrix} &  \gamma_4 &= 
      \frac{1}{\sqrt{2}}\begin{pmatrix}
        0 & 0 & \imath  \\
        0 & 0 & 0 \\
        \imath  & 0 & 0
      \end{pmatrix} \\
      \gamma_5 &= 
      \frac{1}{\sqrt{2}}\begin{pmatrix}
        0 & 0 & 1 \\
        0 & 0 & 0 \\
        -1 & 0 & 0
      \end{pmatrix} &  \gamma_6 &= 
      \frac{1}{\sqrt{2}}\begin{pmatrix}
        0 & 0 & 0 \\
        0 & 0 & \imath  \\
        0 & \imath  & 0
      \end{pmatrix}  & \gamma_7 &= 
      \frac{1}{\sqrt{2}}\begin{pmatrix}
        0 & 0 & 0 \\
        0 & 0 & 1 \\
        0 & -1 & 0
      \end{pmatrix} & \gamma_8 &= 
      \frac{1}{\sqrt{6}}\begin{pmatrix}
        \imath  & 0 & 0 \\
        0 & \imath  & 0 \\
        0 & 0 & -2 \imath 
      \end{pmatrix}
    \end{array}
  \end{equation}
\end{small}
which presents the advantage of being orthonormal $\kappa \left( \lambda_i, \lambda_j
\right) = \delta_{ij}$. In this case the Cartan subalgebra is generated by
$\mathfrak{k} = \braket{\lambda_3, \lambda_8}$.

\section{Characters of affine Lie algebras}
\label{appLie}

In this section we will recall some facts about the partition functions 
and characters of affine Lie algebras. The characters 
of an affine Lie algebra $\hat{\mathfrak{g}}$ are the generating functions 
of the weights multiplicities in a given irreducible representation 
of highest weight $\Lambda$:
\begin{equation}
  ch_{\Lambda} \left(\tau,\nu,u\right) = e^{-2\imath \pi k u} 
  \sum_{\hat{\lambda} \in \text{Rep} \left(\Lambda\right)} \text{dim} V_{\hat{\lambda}} 
  \exp \{ 2\imath \pi \tau n + \sum_i \nu_i \kappa \left(e_i , \hat{\lambda}\right) 
  \}\, ,
\end{equation}
where $\text{dim} V_{\hat{\lambda}}$ is the multiplicity of the 
affine weight $\hat{\lambda}= \left(\lambda,k,n\right)$ 
and $\{ e_i \}$ an orthonormal basis of the root space.
In the framework \textsc{cft} we define slightly different characters, weighted 
by the conformal dimension of the highest weight of the representation:
\begin{equation}
  \chi^{\Lambda} \left(\tau,\nu,u\right) = e^{-2\imath \pi k u} 
  \text{Tr}_{rep \left(\Lambda\right)} \left[ q^{L_0 -\nicefrac{c}{24}}
    e^{2\imath \pi \kappa \left(\nu,\mathcal{J}\right)} \right]
  = e^{2\imath \pi \tau \frac{\kappa \left(\Lambda,\Lambda + 2\rho\right)}{2\left(k
        +\mathfrak{g}^\ast \right)} 
    - \nicefrac{c}{24}} ch_{\Lambda} \left(\tau,\nu,u\right)\, .
\end{equation}
where $\rho = \sum_{\alpha>0} \alpha /2$ and $\mathfrak{g}^\ast$ the dual Coxeter number.
To each affine weight $\hat{\lambda}$ we shall assign a theta-function as
follows:
\begin{equation}
  \Theta_{\hat{\lambda}} \left(\tau, \nu , u\right) =  e^{-2\imath \pi k u} 
  \sum_{\gamma \in \mathbf{M_{\textsc{l}}} + \frac{\lambda}{k}} 
  e^{\imath \pi \tau k \, \kappa \left(\gamma,\gamma\right)} \ 
  e^{2\imath \pi k \kappa \left(\nu, \gamma\right)}
\end{equation}
with $\mathbf{M_{\textsc{l}}}$ the the long roots lattice. 
We can write the affine characters in terms of the theta-function 
with the Weyl-Ka\v c formula:
\begin{equation}
  \chi^{\Lambda} \left( \tau, \nu ,u \right) = \frac{\displaystyle{\sum_{w \in \mathbf{W}}} 
    \epsilon\left(w\right) \Theta_{w\left(\hat{\Lambda} + \hat{\rho}\right)} \left(\tau,\nu,u\right)}
  {\displaystyle{\sum_{w \in \mathbf{W}}} 
    \epsilon\left(w\right) \Theta_{w\left(\hat{\rho}\right)} \left(\tau,\nu,u\right)}\, ,
\end{equation}
$\mathbf{W}$ being the Weyl group of the algebra and $\epsilon \left(w\right)$ the 
parity of the element $w$.

These affine characters are the building blocks of the modular invariant
partition function for the \textsc{wzw} model, since the affine Lie algebra
is the largest chiral symmetry of the theory:
\begin{equation}
  Z = \sum_{\Lambda , \bar \Lambda} M^{\Lambda \bar \Lambda} 
  \chi^{\Lambda} \left(\tau,0,0 \right)
    \bar{\chi}^{\bar \Lambda} \left(\bar{\tau},0,0\right)
\end{equation}
where the sum runs over left and right representations of $\mathfrak g$ with
highest weight $\Lambda$ and $\bar \Lambda$. The representations 
appearing in this partition function are the {\it integrable} ones, 
which are such that:
\begin{equation}
  \text{Rep} \left(\Lambda\right) \text{ integrable } \Longleftrightarrow 
  \frac{2}{\kappa \left(\theta,\theta\right)} \left[ k - \kappa \left(\Lambda,\theta\right) \right] 
  \in \mathbb{N}\, ,
\end{equation}
where $\theta$ is the highest root. The matrix $M^{\Lambda \bar \Lambda}$ 
is such that the partition function of $\mathfrak{g}_k$ is modular invariant; 
at least, the diagonal $\delta_{\Lambda , \bar \Lambda}$
exists since the characters form an unitary representation of the 
modular group. 

In the heterotic strings, the worldsheet has a local $N=\left(1,0\right)$ local 
supersymmetry so the left algebra is lifted to a super-affine 
Lie algebra. However the characters can be decoupled as characters 
of the bosonic algebra times characters of free fermions:
\begin{equation}
  Z \oao{a}{b} = \sum_{\Lambda , \bar \Lambda} M^{\Lambda \bar \Lambda} 
  \chi^{\Lambda} \left(\tau \right) \ 
  \left( \frac{\vartheta\oao{a}{b} \left(\tau \right)}{\eta \left(\tau \right)} \right)^{
    \mathrm{dim} \left( \mathfrak g \right) /2} 
  \bar{\chi}^{\bar \Lambda} \
\end{equation}
where $\left(a,b\right)$ are the spin structures of the worldsheet fermions.

The characters of the affine algebras can be decomposed according to 
the generalized parafermionic decomposition, by factorizing 
the abelian subalgebra of the Cartan torus. 
For example, we can decompose the left supersymmetric $\mathfrak{g}_k$ characters
in terms of characters of the supersymmetric coset, given 
by the following branching relation (see~\cite{Kazama:1989qp}):
\begin{equation}
  \chi^{\Lambda}  \ 
  \left(
    \frac{\vartheta \oao{a}{b}}{\eta}\right)^{\mathrm{dim} \left(\mathfrak j \right) /2}  = 
  \sum_{\lambda \ \mathrm{mod} \ \left(k + g^* \right) \mathbf{M_\textsc{l}}} 
  \mathcal{P}^{\Lambda}_{\lambda} \oao{a}{b}  
  \frac{\Theta_{\lambda,k+g^*}}{\eta^{\mathrm{dim} \left(\mathfrak k \right)}}
\end{equation}
in terms of the theta-functions associated to $\mathfrak g_k$.

\subsection{The example of $SU\left(3\right)$}

In an orthonormal basis, the simple roots of $SU\left(3\right)$ are: 
\begin{align}
  \alpha_1 = \left(\sqrt{2},0\right)\ , &&  \alpha_2 = \left(-
  \nicefrac{1}{\sqrt{2}},\sqrt{\nicefrac{3}{2}}\right).
\end{align}
The dual basis of the fundamental weights, defined by 
$\left(\lambda_{f}^i , \alpha_j \right) = \delta^{i}_{\, j}$ is given by:
\begin{align}
  \lambda_{f}^1 = \left(\nicefrac{1}{\sqrt{2}} , \nicefrac{1}{\sqrt{6}} \right)
  \ , &&
  \lambda_{f}^2 = \left(0,\sqrt{\nicefrac{2}{3}}\right).
\end{align}
As they should the simple roots belong to the weight lattice:
\begin{align}
  \alpha_1 = 2\lambda_{f}^1 - \lambda_{f}^2 \ , &&
  \alpha_2 = 2\lambda_{f}^2 - \lambda_{f}^1.
\end{align}
The theta function of the $\widehat{\mathfrak{su}}_3$ affine algebra reads, 
for a generic weight $\lambda = m_i \lambda_{f}^i$:
\begin{equation}
  \Theta_{\lambda,k} = \sum_{\gamma \in M} 
  q^{\frac{k}{2} \| \gamma + \frac{\lambda}{k} \|^2} = 
  \sum_{n^1, n^2} 
  q^{\frac{k}{2} \| n^1 \alpha_1 + n^2 \alpha_2 
    + \frac{m_1 \Lambda^1 + m_2 \Lambda^2}{k} \|^2}
\end{equation}
So the vector appearing in the theta function is: 
\begin{equation}
  \left\{ \sqrt{k}\left(2n^1 -n^2\right) + \frac{m_1}{\sqrt{k}} \right\} \frac{e_1}{2} 
  + \left\{  \sqrt{k} n^2 + \frac{m_1+2m_2}{3\sqrt{k}} \right\}
  \frac{\sqrt{3}}{2}\, e_2
\end{equation}

\subsection{Modular transformations}
We have the following modular transformations for the theta-functions:
\begin{equation}
\Theta_{\lambda,k} \left(-1/\tau \right) = \left(-i\tau\right)^{\mathrm{dim} \left(\mathfrak k \right) /2}
\left| \frac{\mathbf{M}^\ast}{k\mathbf{M_\textsc{l}}} \right|^{-1/2} \sum_{\mu \in 
\mathbf{M}^\ast \mod k\mathbf{M_\textsc{l}}}
e^{2i\pi \left(\lambda, \mu\right) /k} \Theta_{\mu,k} \left(\tau \right),
\end{equation}
where $\mathbf{M}^\ast$ is the lattice dual to $\mathbf{M_\textsc{l}}$, 
$|\mathbf{M_\textsc{l}}|$ is the size of the 
basic cell of $\mathbf{M_\textsc{l}}$ and for the affine characters:
\begin{equation}
\chi^{\Lambda} \left(-1/\tau \right) = \left| \frac{\mathbf{M}^\ast}{\left(k+g^\ast\right)
\mathbf{M_\textsc{l}}} 
\right|^{-1/2} i^{|\Delta_+|} \sum_{\Lambda '} \sum_{w \in W} 
\epsilon \left(w\right) \, e^{\frac{2i\pi}{k+g^\ast} \left(\Lambda + \rho \right)w\left(\Lambda' + \rho\right)}
\, \chi^{\Lambda'} \left(\tau \right)
\end{equation}
In this formula, $|\Delta_+|$ is the number of positive roots.
From these two formulas we deduce the modular transformation 
of the characters of the super-coset under $\tau  \to -1/\tau$:
\begin{equation}
\mathcal{C}^{\Lambda}_{\lambda} \oao{a}{b} \left(-1/\tau \right) 
= e^{\frac{i\pi}{4} ab \dim \left(\mathfrak j\right)} \,
i^{|\Delta_+|} 
\sum_{\mu \in 
\mathbf{M}^\ast \, \mathrm{mod}\, k\mathbf{M_\textsc{l}}} 
e^{2i\pi \frac{\left(\lambda ,\mu\right)}{k+g^\ast}}\,
 \sum_{\Lambda '} \sum_{w \in W} 
\epsilon \left(w\right) \, e^{\frac{2i\pi}{k+g^\ast} \left(\Lambda + \rho \right)w\left(\Lambda' + \rho\right)}
\, \mathcal{C}^{\Lambda'}_{\mu} \oao{b}{-a} \left( \tau \right)
\end{equation}

\subsection{Fermionic characters}

For an even number of fermions it is possible to express the characters 
in terms of representations of the $SO\left(2n\right)_1$ affine algebra. 
The characters are labelled by $s=\left(0,1,2,3\right)$ for the 
trivial, spinor, vector and conjugate spinor representations:
\begin{eqnarray}
\Xi_{2n}^0 = &\frac{1}{2\eta^n} \left[ \theta \oao{0}{0}^n + 
\theta \oao{0}{1}^n \right] & \text{trivial}\nonumber\\
\Xi_{2n}^2 = &\frac{1}{2\eta^n} \left[ \theta \oao{0}{0}^n - 
\theta \oao{0}{1}^n \right] & \text{vector}\nonumber\\
\Xi_{2n}^1 = &\frac{1}{2\eta^n} \left[ \theta \oao{1}{0}^n + \imath^{-n}
\theta \oao{1}{1}^n \right] & \text{spinor}\nonumber\\
\Xi_{2n}^3 = &\frac{1}{2\eta^n} \left[ \theta \oao{1}{0}^n - \imath^{-n}
\theta \oao{1}{1}^n \right] & \text{conjugate spinor}
\end{eqnarray}
Their modular matrices are:
\begin{equation}
  T = e^{-\imath n \pi/12} 
  \begin{pmatrix}
    1&0&0&0\\
    0&-1&0&0\\
    0&0&e^{\imath n \pi/4}&0\\
    0&0&0&e^{\imath n \pi/4}
  \end{pmatrix}
\end{equation}
and
\begin{equation}
  S = \frac{1}{2} 
  \begin{pmatrix}
    1&1&1&1\\
    1&1&-1&-1\\
    1&-1&\imath^{-n}&-\imath^{-n}
    \\1&-1&-\imath^{-n}& \imath^{-n}
  \end{pmatrix}
\end{equation}


\bibliography{Biblia}

\end{document}